\documentclass[twocolumn,aps,pra]{revtex4}

\usepackage{mathrsfs}
\usepackage{amsfonts}
\usepackage{amsmath}

\newcommand{\be}{\begin{equation}}
\newcommand{\ee}{\end{equation}}
\newcommand{\ba}{\begin{array}}
\newcommand{\ea}{\end{array}}
\newcommand{\bw}{\begin{widetext}}
\newcommand{\ew}{\end{widetext}}

\newcommand{\ra}{\rangle}
\newcommand{\la}{\langle}
\newcommand{\pp}{\partial}
\newcommand{\ov}{\overline}

\newcommand{\wh}{\widehat}
\newcommand{\ww}{\widetilde}

\newcommand{\bk}{{\bf k}}
\newcommand{\bp}{{\bf p}}

\newcommand{\bq}{{\bf q}}

\newcommand{\bx}{{\bf x}}

\newcommand{\A}{{\mathscr{A}}}

\newcommand{\E}{{\cal E}}

\newcommand{\LL}{\mathcal{L}}
\newcommand{\N}{{\mathscr{N}}}

\newcommand{\WW}{{\mathscr{W}}}

\newcommand{\VV}{{\mathscr{V}}}

\begin{document}

 \title{Symmetries with the same forms as gauge symmetries in the electroweak theory
 }

\author{Wen-ge Wang}
\affiliation{
 Department of Modern Physics, University of Science and Technology of China,
 Hefei 230026, China
 }

 \date{\today}

\begin{abstract}
 Within the electroweak theory, it is shown that the form of the total Lagrangian is invariant, under
 local phase changes of the basis states for leptons and
 under local changes of the mathematical spaces employed for the description of
 left-handed spinor states of leptons.
 In doing this, a contribution from vacuum fluctuations of the leptonic fields, 
 which causes no experimentally-observable effect, is added to the total connection field.
 Accompanying the above-mentioned changes of basis states, the leptonic and connection fields
 are found to undergo changes whose forms are similar to $U(1)$ and $SU(2)$ gauge transformations, respectively.
 These results suggest a simple physical interpretation to gauge symmetries in the electroweak theory.
\end{abstract}

 \maketitle


\section{Introduction}

 Gauge symmetry and Lorentz symmetry
 are the most important symmetries in modern physics,
 particularly, in the establishment of the standard model (SM) \cite{Weinberg-book,Peskin,Dai-YB}.
 The physical origin of the Lorentz symmetry is quite clear,
 which is based on the fact that
 no inertial frame of reference has been found physically superior to any other one.
 In contrast, physical origin of the gauge symmetry is much less clear.
 Further understanding of the latter should be useful
 in the study of various topics, such as going beyond the SM,
 which has attracted lots of attention in recent years
 (see, e.g., Refs.\cite{GGS99,KO01,Polch-book,Ross84,Raby93,book-fundam-QFT,GN03,SV06}).

 One interesting question is whether gauge transformations of leptonic fields, or part of them,
 may be related to changes of basis states that are employed in the description of leptonic states.
 As an example, one may consider the quantum electrodynamics (QED),
 in which a gauge transformation of the electronic field $\psi_e(x)$ is written as
 $\psi_e(x) \to  e^{-i\varphi(x)} \psi_e(x)$.
 This transformation has a form that
 is similar to the change of a single-particle wave function $\phi(x) = \la x|\phi\ra$
 under the following change of the basis states $|x\ra$,
 namely,  $|x\ra \to e^{i\varphi(x)}|x\ra$, resulting in $\phi(x) \to e^{-i\varphi(x)}\phi(x)$.
 Although this similarity is instructive, to take it in a serious way, an immediate problem is 
 that the field $\psi_e(x)$ is much more complex than a single-particle wave function. 
 Moreover, to get deeper understandings for gauge symmetries
 along this line of approach,  several further questions should be answered. 
 Particularly,  (i) whether gauge transformations of bosonic fields may be understood
 in some related way;
 and (ii) whether this approach could be useful for more complicated gauge transformations, such as
 $SU(2)$ transformations in the Glashow-Weinberg-Salam (GWS) electroweak theory
 \cite{Weinberg-book,Peskin,Dai-YB},
 which involve rotations in an intrinsic degree of freedom of leptons.

 In this paper, we show that the above questions have positive answers within the electroweak theory.
 Particularly, it is shown that, under certain local changes of the basis states for leptons, 
 the total Lagrangian possesses symmetries whose forms are similar to the $U(1)$ and $SU(2)$ 
 gauge symmetries.
 We are to focus on the first generation of leptons,
 because other generations can be treated in the same way.

 In order to study changes of leptonic spinor bases and spaces,
 an appropriate framework is supplied by the spinor theory
 \cite{Penrose-book,Kim-group,CM-book,Corson,pra16-commu},
 which is based on the so-called $SL(2,C)$ group,
 a covering group of the proper, orthochronous Lorentz group.
 Although the relationship between Dirac spinors and two-component Weyl spinors 
 in the spinor theory is
 well known, detailed properties of Weyl spinors are not widely discussed in physics.
 Moreover, the ordinary notation used in the spinor theory
 is not always convenient when discussing quantum states.
 For these reasons,  before discussing the physics,  in Sec.\ref{sect-2-spinor}
 we recall basic results of the spinor theory
 and write them in an abstract notation that uses Dirac's kets and bras.

 We found that the total Lagrangian in the GWS theory may have the above-mentioned symmetries,
 if a contribution from vacuum fluctuations of the leptonic fields,
 which causes no experimentally-observable effect, is added to the total connection field.
 This small modification to the GWS theory is discussed in Sec.\ref{sect-vacuum-contribution}.
 In Sec.\ref{sect-QED}, we discuss local phase changes of basis states for leptons,
 which leads to results that have forms similar to $U(1)$ gauge transformations.
 Then, in Sec.\ref{sect-SU2}, we discuss certain changes of the spinor spaces
 for left-handed (LH) spinor states of leptons.
 It is shown that this leads to changes of fields that have forms
 similar to $SU(2)$ gauge transformations.
 Finally, conclusions and discussions are given in Sec.\ref{sect-conclusion}.

\section{Spinors and Vectors in an abstract notation}\label{sect-2-spinor}

 In this section, we recall basic properties of spinors \cite{Penrose-book,CM-book,Corson,Kim-group,pra16-commu},
 which are useful in the discussions to be given in later sections.
 The theory of spinors usually employs a notation, in which spinors are described by their components.
 But, in the study of changes of spinor bases, it is more convenient
 to write spinors in the abstract notation of Dirac's kets and bras
 as done in Ref.\cite{pra16-commu}.
 Specifically, we discuss basic properties of two-component Weyl spinors in the abstract notation
 in Sec.\ref{sect-recall-spinor},
 and discuss the abstract notation for stationary solutions of the Dirac equation in Sec.\ref{sect-Dirac-spinor}.
 We recall basic properties of four-component vectors
 in Sec.\ref{sect-recall-vector} and discuss their abstract expressions
 in Sec.\ref{sect-vector-abstract}.

 Two conventions are to be followed in this paper, unless otherwise stated.
 The first one is that each pair of repeated label in a product implies summation over the label.
 The second one, which is used in the spinor theory,
 is that an overline indicates complex conjugate;
 for this reason, we do not write, say, $U^\dag \gamma^0$ as $\ov U$.

\subsection{Basic properties of two-component spinors}\label{sect-recall-spinor}

 In this section, we give a brief discussion for Weyl spinors in
 Dirac's abstract notation \cite{cite-pre16}.
 In the spinor theory, there are two smallest nontrivial representation spaces of the $SL(2,C)$ group,
 which are spanned by two types of Weyl spinors, with the relationship of complex conjugation.
 We use $\WW$ to denote one of these two spaces.

 In terms of components, a Weyl spinor in $\WW$ is written as, say,
 $\kappa_A$ with an index $A=0,1$.
 In the abstract notation, a basis in the space $\WW$ is written as $|S^{A}\ra $
 and the above spinor is expanded as
\begin{equation}\label{|kappa>}
  |\kappa\ra = \kappa_A|S^{A}\ra,
\end{equation}
 where a summation over $A$ is implied by a convention discussed above.
 One may introduce a space that is dual to $\WW$, composed of bras with a basis written as $\la S^A|$.
 In order to construct a product that is a scalar under $SL(2,C)$ transformations,
 the bra dual to the ket $|\kappa\ra$ should be written as
\begin{equation}\label{<kappa|}
  \la \kappa | =  \la S^{A}|  \kappa_A,
\end{equation}
 which has the same components as $|\kappa\ra$ in Eq.(\ref{|kappa>}),
 but not their complex conjugates.
 (See Appendix \ref{sect-SL2C-transf} for a brief discussion of some basic properties
 of $SL(2,C)$ transformations.)

 Scalar products of the basis spinors satisfy
\begin{equation}\label{SA-SB}
  \la S^{A}|S^{B}\ra = \epsilon^{A B},
\end{equation}
 where
\begin{equation}\label{epsilon}
 \epsilon^{AB} = \left( \begin{array}{cc} 0 & 1 \\ -1 & 0 \end{array} \right).
\end{equation}
 It proves convenient to introduce another matrix $\epsilon_{AB}$, which has the same
 elements as $\epsilon^{AB}$.
 These two matrices can be used to raise and lower indexes of components,  say,
\begin{equation}\label{A-raise}
  \kappa^A = \epsilon^{AB} \kappa_B, \quad \kappa_A = \kappa^B \epsilon_{BA},
\end{equation}
 as well as for the basis spinors, namely, $|S_A\ra = |S^B\ra \epsilon_{BA}$
 and $|S^A\ra = \epsilon^{AB}|S_B\ra $.
 It is not difficult to verify that (i) $\la S_{A}|S_{B}\ra = \epsilon_{A B}$; (ii)
\begin{equation}\label{f-AB}
  {f_{\ldots}^{\ \ \ A}\ (g)^{  \cdots }}_{ A } = - {f_{\ldots A}\ (g)^{\cdots A}};
\end{equation}
 and (iii) the symbols $\epsilon_C^{\ \ A} =\epsilon^{BA} \epsilon_{BC}$ and
 $\epsilon^A_{\ \ C} =\epsilon^{AB} \epsilon_{BC}$ satisfy the relation
\begin{gather}\label{eps-delta}
 \epsilon_C^{\ \ A} = -\epsilon^A_{\ \ C} = \delta^A_C,
\end{gather}
 where $\delta^A_B=1$ for $A=B$ and $\delta^A_B=0$ for $A \ne B$.

 The scalar product of two generic spinors $|\chi\ra$ and $|\kappa\ra$, written as
 $\la\chi|\kappa\ra$, has the expression
\begin{equation}\label{<chi|kappa>}
 \la\chi|\kappa\ra = \chi_A \kappa^A.
\end{equation}
 The anti-symmetry of $\epsilon_{AB}$ implies that
\begin{equation}\label{ck=-kc}
  \la \chi |\kappa\ra = -\la\kappa |\chi\ra
\end{equation}
 and, as a consequence, $\la \kappa |\kappa\ra =0$ for all $|\kappa\ra$.
 Moreover, we note the following properties:
 (i) The identity operator $I_{\WW}$ in the space $\WW$ can be written as
\begin{eqnarray}\label{I}
 I_{\WW} = |S^{A}\ra \la S_{A}|,
\end{eqnarray}
 satisfying $I_{\WW}|\kappa\ra =|\kappa\ra $ for all $|\kappa\ra \in \WW$;
 and (ii) the components of $|\kappa\ra$
 have the following expressions,
\begin{equation}\label{kappa-A}
  \kappa^A = \la S^{A}|\kappa\ra, \quad \kappa_A = \la S_{A}|\kappa\ra.
\end{equation}

 An operation of complex conjugation can be introduced,
 converting $\WW$ to a space denoted by $\ov\WW$,
 which is the other representation space of the $SL(2,C)$ group mentioned
 in the beginning of this section.
 This  operation changes spinors $|\kappa\ra$ in $ \WW$
 to spinors in $\ov\WW$, written as $|\ov\kappa\ra$.
 Corresponding to the basis $|S_{A}\ra \in \WW$, the space $\ov \WW$ has a basis
 denoted by $|\ov S_{A'}\ra$ with a primed label $A' = 0', 1'$.
 In the basis of $|\ov S_{A'}\ra$, $|\ov\kappa\ra$ is written as
\begin{equation}\label{ov-kappa}
  |\ov\kappa \ra = {\ov\kappa}_{A'}|\ov S^{A'}\ra,
\end{equation}
 where
\begin{equation}\label{}
  \ov\kappa^{A'} := (\kappa^A)^* .
\end{equation}
 Similar to the $\epsilon$ matrices discussed above,
 one introduces matrices $\epsilon^{A'B'}$ and $\epsilon_{A'B'}$,
 which have the same elements as $\epsilon^{AB}$ and are used to raise and lower primed labels.
 The identity operator in the space $\ov\WW$, denoted by $I_{\ov\WW}$
 has the form $I_{\ov\WW} = |\ov S^{A'}\ra \la \ov S_{A'}|$.
 When a spinor $\kappa^A$ is transformed by an $SL(2,C)$ matrix,
 the spinor $\ov\kappa^{A'}$ is transformed by the complex-conjugate matrix
 (see Appendix \ref{sect-SL2C-transf}).

\subsection{Dirac spinors in the abstract notation}\label{sect-Dirac-spinor}

 In this section, we briefly discuss some properties of Dirac spinors written
 in the abstract notation of Dirac's kets and bras  \cite{cite-pre16}.
 The Dirac equation for a free electron with a mass $m$
 has two solutions labelled by a Lorentz invariant index $r=0,1$, i.e.,
 $U^r(\bp) e^{-ipx}$ with a three-momentum $\bp$ and a four-momentum $p$,
 where   $x= x^\mu$ and $p = p^\mu = (p^0,\bp)$ with $\mu=0,1,2,3$,
 satisfying $p^\mu p_\mu =m^2$ with  $p^0>0$.
 Here, $U^r(\bp)$ are four-component Dirac spinors  satisfying
\begin{equation}\label{stat-DE}
  (\gamma^\mu p_\mu -m) U^r(\bp)=0.
\end{equation}
 In the chiral representation of the $\gamma^\mu$-matrices,
 a four-component Dirac spinor is decomposed into two two-component Weyl spinors,
 the LH part and the right-handed (RH) part \cite{Penrose-book,Corson,CM-book,Peskin}.
 Specifically, the spinor $U^r(\bp)$ is written as
\begin{gather} \label{Up-uv-main}
 U^r(\bp) = \frac{1}{\sqrt 2} \left( \begin{array}{c} u^{r,A}(\bp) \\ \ov v_{B'}^r(\bp) \end{array} \right).
\end{gather}

 In labels for two-component spinors, the $\gamma^\mu$-matrices are written as
\begin{gather} \label{gamma-mu}
 \gamma^\mu = \left( \begin{array}{cc} 0 & \sigma^{\mu AB'}
 \\ \ov\sigma^{\mu}_{A'B} & 0\end{array}\right),
\end{gather}
 where $\sigma^{\mu AB'}$ are the so-called Enfeld-van der Waerden symbols, in short \emph{EW-symbols}
 \cite{Penrose-book,Kim-group,CM-book,Corson,pra16-commu}.
 Here, $\ov\sigma^{\mu}_{A'B}$ indicates the complex conjugate of $\sigma^{\mu}_{AB'}$, namely,
 $\ov\sigma^{\mu}_{A'B} =(\sigma^{\mu}_{AB'})^*$.
 An often-used set of explicit expressions for these symbols is given below,
\begin{eqnarray}\notag
 \sigma^{0AB'} = \left(\begin{array}{cc} 1 & 0 \\ 0 & 1 \\ \end{array} \right),
 \sigma^{1AB'}  =  \left(\begin{array}{cc} 0 & 1 \\ 1 & 0 \\ \end{array} \right),
 \\ \sigma^{2AB'} = \left(\begin{array}{cc} 0 & -i \\ i & 0 \\ \end{array} \right),
 \sigma^{3AB'}  =  \left(\begin{array}{cc} 1 & 0 \\ 0 & -1 \\ \end{array} \right).
 \label{sigma^AB}
\end{eqnarray}
 The stationary Dirac equation (\ref{stat-DE}) is, then, split into
 two equivalent subequations, namely,
\begin{subequations}\label{Deq-2p}
\begin{eqnarray}
 \label{sv-u-1m}   p_\mu \sigma^{\mu AB'} \ov v_{B'}^r(\bp) - mu^{r,A}(\bp) =0,
 \\ \label{sv-u-2m}  p_\mu  \ov\sigma^{\mu}_{B'A} u^{r,A}(\bp) - m\ov v_{B'}^r(\bp) =0.
\end{eqnarray}
\end{subequations}

 Correspondingly, a solution for a free positron with a four-momentum $p$
 ($p^0>0$) is usually written as $V^r(\bp) e^{ipx}$, satisfying
\begin{equation}\label{stat-DE-V}
  (\gamma^\mu p_\mu +m) V^r(\bp)=0,
\end{equation}
 where
\begin{gather} \label{Vp-uv-main}
 V^r(\bp) = \frac{1}{\sqrt 2} \left( \begin{array}{c} u^{r,A}(\bp) \\ -\ov v_{B'}^r(\bp) \end{array} \right).
\end{gather}

 In the abstract notation, the Weyl spinors $u^{r,A}(\bp)$ and $ \ov v_{B'}^r(\bp)$ are written as
\begin{subequations} \label{|uv>}
\begin{gather}\label{|u>-uA}
 |u^r(\bp)\ra = u^{r}_{A}(\bp)|S^A\ra = -u^{r,A}(\bp)|S_A\ra,
 \\ |\ov v^r(\bp)\ra = \ov v^{r}_{B'}(\bp)|S^{B'}\ra.
\end{gather}
\end{subequations}
They satisfy the following relations,
\begin{gather}\label{uu-eps}
 \la u^r(\bp) |u^{s}(\bp)\ra =
  \la \ov v^{r}(\bp)| \ov v^{s}(\bp)\ra = \epsilon^{rs},
 \\  \la v^r(\bp)| u^r(\bp)\ra = \delta^{rs}. \label{ip-vu}
\end{gather}
 Then, the Dirac spinors $U^r(\bp)$ and $V^r(\bp)$ are written as
\begin{subequations} \label{|U-Vp>}
\begin{gather}
 |U^r(\bp)\ra = \frac{1}{\sqrt 2} \left( \begin{array}{c} |u^r(\bp) \ra \\ |\ov v^r(\bp)\ra \end{array} \right),
 \\ |V^r(\bp)\ra = \frac{1}{\sqrt 2} \left( \begin{array}{c} |u^r(\bp) \ra \\ -|\ov v^r(\bp)\ra \end{array} \right).
\end{gather}
\end{subequations}
 To be consistent with the expression of bra in Eq.(\ref{<kappa|})
 for two-component spinors,
 the bras corresponding to the above two kets should be written as
\begin{subequations}\label{<U-Vp|}
\begin{gather}
 \la U^r(\bp)| = \frac{1}{\sqrt 2} \left(  \la u^r(\bp)| , \la\ov v^r(\bp)| \right),
 \\ \ \la V^r(\bp)| =  \frac{1}{\sqrt 2} \left( \la u^r(\bp)| , -\la \ov v^r(\bp)| \right),
\end{gather}
\end{subequations}
 without taking complex conjugate for the two-component spinors.
 Direct derivation shows that
\begin{gather}\label{<Ur|Us>}
 \la U^r(\bp)|U^s(\bp)\ra = \la V^r(\bp)|V^s(\bp)\ra = \epsilon^{rs}.
\end{gather}
 Meanwhile, the complex conjugates of, say,
 $|U^r(\bp)\ra$ and $\la U^r(\bp)|$ are written as
\begin{subequations}\label{|wh-U>}
\begin{gather}
 |\ov U^r(\bp)\ra =\frac{1}{\sqrt 2} \left( \begin{array}{c} |\ov u^r(\bp) \ra \\ |v^r(\bp)\ra \end{array} \right),
 \\  \la \ov U^r(\bp)| =\frac{1}{\sqrt 2} \left( \la\ov u^r(\bp)| , \la v^r(\bp)| \right).
\end{gather}
\end{subequations}

 Although a product $\la U'(\bp)|U(\bp)\ra$ is a Lorentz scalar, it is not an inner product,
 because Eq.(\ref{<Ur|Us>}) gives that $\la U(\bp)|U(\bp)\ra =0$.
 In order to get the well-known inner product used in QED written
 in the abstract notation, the matrix $\gamma_0$ that is used in ordinary
 expressions such as $U^\dag \gamma_0$
 should be replaced by the following matrix $\gamma_c$  \cite{pra16-commu},
\begin{equation} \label{gamma-c}
 \gamma_c =  \left( \begin{array}{cc} 0 & -1 \\ 1 & 0 \end{array} \right).
\end{equation}
 Then, the inner product is written in a form of, say, $\la \wh U^{r}(\bp)|U^{s}(\bp)\ra$,
 where
\begin{equation}\label{ov-X}
  \la\wh U^{r}(\bp)| := \la \ov U^{r}(\bp)|  \gamma_c =  (\la v^{r}(\bp)|,-\la \ov u^{r}(\bp)|).
\end{equation}
 It is straightforward to verify the following relations,
\begin{subequations} \label{|UV-IP>}
\begin{gather}\label{UrUs-cc}  
 \la \wh U^{r}(\bp)|U^{s}(\bp)\ra = \delta^{rs},
 \quad \la {\wh{V}^{r}}(\bp)|V^{s}(\bp)\ra = \delta^{rs},
 \\ \la {\wh{U}^{r}}(\bp)|V^{s}(\bp)\ra = 0, \quad
  \la {\wh{V}^{r}}(\bp)|U^{s}(\bp)\ra = 0.
\end{gather}
\end{subequations}

 \subsection{Basic properties of four-component vectors}\label{sect-recall-vector}

 In this section, we recall basic properties of four-component vectors
 as special cases of spinors \cite{Penrose-book,Corson,CM-book}.
 We use the ordinary notation in this section and
 will discuss the abstract notation in the next section.

 A basic point is a one-to-one mapping between a direct-product
 space $\WW \otimes\ov\WW$ and a four-dimensional vector space denote by $\VV$,
 which is given by the EW-symbols discussed above.
 For example,  by the following relation, a spinor $\phi_{AB'}$ in the space
 $\WW \otimes\ov\WW$ can be mapped to a vector in the space $\VV$, denoted by  $K^\mu$,
\begin{equation}\label{map-WW-V}
  K^\mu = \sigma^{\mu AB'} \phi_{AB'}.
\end{equation}
 In the space $\VV$, of particular importance is a symbol denoted by $g^{\mu\nu}$, defined
 by the following relation with the $\epsilon$-symbols discussed previously,
\begin{equation}\label{g-sig}
 g^{\mu\nu} =  \sigma^{\mu AB'} \sigma^{\nu CD'} \epsilon_{AC} \epsilon_{B'D'}.
\end{equation}
 One can introduce a lower-indexed symbol $g_{\mu\nu}$,
 which has the same matrix elements as $g^{\mu\nu}$, namely, $[g^{\mu\nu}] = [g_{\mu\nu}]$.
 These two symbols $g$, like the symbols $\epsilon$ in the space $\WW$,
 can be used to raise and lower indexes, e.g.,
\begin{equation}\label{mu-raise}
  K_\mu =  K^\nu g_{\nu \mu}, \quad K^\mu = g^{\mu\nu} K_\nu.
\end{equation}
 Making use of the anti-symmetry of the symbols $\epsilon$,
 it is easy to verify that $g^{\mu\nu}$ is symmetric, i.e.,
\begin{equation}\label{g-sym}
 g^{\mu\nu} = g^{\nu\mu}.
\end{equation}
 Due to this symmetry, the upper/lower positions of repeated
 indexes ($\mu$) are exchangeable, namely
\begin{equation}\label{f-munu}
  {F_{\ldots}^{\ \ \ \mu}(f)^{  \cdots }}_{\mu } = {F_{\ldots \mu}(f)^{\cdots \mu}}.
\end{equation}

 The EW-symbols have the following properties,
\begin{equation}\label{st-delta}
 \sigma^{AB'}_\mu \sigma_{CD'}^\mu = \delta^{AB'}_{CD'}, \quad
 \sigma_{AB'}^\mu \sigma^{AB'}_\nu  = \delta_\nu^\mu,
\end{equation}
 where $\delta_\nu^\mu = 1$ for $\mu =\nu$ and $\delta_\nu^\mu = 0$ for $\mu \ne\nu$,
 and $ \delta^{AB'}_{CD'} :=  \delta^{A}_{C} \delta^{B'}_{D'}$.
 Making use of the relations in Eq.(\ref{st-delta}), it is not difficult to check that the map
 from $\WW\otimes\ov\WW$ to $\VV$ given in Eq.(\ref{map-WW-V}) is reversible.
 Moreover, noting Eq.(\ref{eps-delta}), one finds that
\begin{gather}\label{ss-ee-2}
 \sigma_{\mu AB'} \sigma_{CD'}^\mu = \epsilon_{AC} \epsilon_{B'D'}.
\end{gather}
 Then, substituting the definition of $g^{\mu\nu}$ in Eq.(\ref{g-sig})
 into the product $g^{\mu\nu} g_{\nu \lambda}$, after simple algebra, one gets that
\begin{gather} \label{ggd}
 g^{\mu\nu} g_{\nu \lambda} = g^\mu_{\ \ \lambda} = g^{\ \ \mu}_{\lambda} = \delta^\mu_\lambda .
\end{gather}

 When a $SL(2,C)$ transformation is carried out on the space $\WW$,
 a related transformation should be applied to the space $\VV$.
 Requiring invariance of the EW-symbols,
 transformations on the space $\VV$ can be fixed
 and turn out to constitute a (restricted) Lorentz group
 (Appendix \ref{sect-SL2C-transf}).
 Therefore, the space $\VV$ is a four-component vector space.
 In fact, substituting the explicit expressions of the EW-symbols in Eq.(\ref{sigma^AB})
 into Eq.(\ref{g-sig}), one gets
\begin{eqnarray}
 g^{\mu\nu} = \sigma^{\mu}_{A'B} \sigma^{\nu A'B}
 =\left(\begin{array}{cccc} 1 & 0 & 0 & 0 \\ 0 & -1 & 0 & 0 \\ 0&0 & -1 &0
 \\ 0 &0 &0 & -1 \end{array} \right),
\end{eqnarray}
 which is just the Minkovski's metric.
 Furthermore,  for two arbitrary vectors $J^\mu$ and $K^\mu$ in the space $\VV$,
 one can show that both of the two products given below,
\begin{gather}\label{Kmu-Jmu}
  J_\nu K^\nu = J^\mu g_{\mu\nu} K^\nu \ \
 \& \ \  J^*_\nu K^\nu = J^{\mu *} g_{\mu\nu} K^\nu,
\end{gather}
 are scalars under Lorentz transformations (Appendix \ref{sect-SL2C-transf}),
 where the second product is the one used in QED.

 \subsection{Abstract notation for four-component vectors}\label{sect-vector-abstract}

 In the abstract notation,
 a basis in the space $\VV$ is written as $|T_\mu\ra $.
 The index of the basis can be raised by $g^{\mu\nu}$,
 i.e., $|T^\mu\ra = g^{\mu\nu} |T_\nu\ra $,  and similarly $ |T_{\mu}\ra = g_{\mu \nu}|T^\nu\ra $.
 A generic four-component vector $|K\ra$ in the space $\VV$ is expanded as
\begin{equation}\label{}
 |K\ra = K_\mu |T^\mu\ra = K^\mu |T_\mu\ra.
\end{equation}
 In consistency with the expression of bra in Eq.(\ref{<kappa|}) for two-component spinors,
 the bra corresponding to $|K\ra$ is written as
\begin{gather}\label{<K|}
 \la K| = \la T_\mu| K^\mu .
\end{gather}
 Requiring that
\begin{eqnarray}\label{Tmu-Tnu}
  \la T_\mu|T_\nu\ra =  g_{\mu\nu},
\end{eqnarray}
 it is easy to check that the scalar product $J_\nu K^\nu$ is written as
 $\la J|K\ra$, namely, $\la J|K\ra = J_\nu K^\nu$,
 similar to the case of two-component spinors in Eq.(\ref{SA-SB}).

 It is not difficult to verify the following properties.
 (i) Making use of Eq.(\ref{ggd}), one finds that the identity operator
 in the space $\VV$ can be written as
\begin{equation}\label{I-vector}
 I_\VV = |T_\mu\ra \la T^\mu | = |T^\mu\ra \la T_\mu|.
\end{equation}
 (ii) The components $K^\mu$ and $K_\mu$ have the following expressions,
\begin{equation}\label{K-mu}
  K^\mu = \la T^\mu |K\ra , \quad K_\mu = \la T_\mu |K\ra .
\end{equation}
 (iii) The symmetry of $g^{\mu\nu}$ implies that $\la T_\mu|T_\nu\ra = \la T_\nu|T_\mu\ra$,
 as a result,
\begin{eqnarray}\label{<K|J>=<J|K>}
  \la K|J\ra = \la J|K\ra.
\end{eqnarray}

 Basis spinors in the space $\WW \otimes\ov\WW$ can be written as
 $|S_{AB'}\ra := |S_A\ra | \ov S_{B'}\ra $, or
 $|S_{B'A}\ra := | \ov S_{B'}\ra |S_A\ra $.
 Due to the generic connection between spin and statistics \cite{Pauli40,SW64},
 we assume anticommutability for the order of the kets (similar for bras), that is
\begin{equation}\label{SAB-commu}
  |S_A\ra | \ov S_{B'}\ra =  - | \ov S_{B'}\ra |S_A\ra \quad \forall A,B'.
\end{equation}
 For basis in the space dual to $\WW \otimes\ov\WW$,
 we write, say, $\la S_{B'A}| := \la \ov S_{B'}| \la S_A|$.

 Since $\ov{|S_{AB'}\ra} =|S_{A'B}\ra= -|S_{BA'}\ra$, the operation of complex conjugation
 maps the space $\VV$ into itself.
 We use $\ov{|T_\mu\ra }$ to denote  the complex conjugate of $|T_\mu\ra $.
 Since $\ov{|T_\mu\ra }$ and $|T_\mu\ra $ lie in the same space,
 it is unnecessary to introduce any change to the label $\mu$.
 Hence, $\ov{|T_\mu\ra }$ can be written as $|\ov T_\mu\ra $ with the
 label $\mu$ unchanged.
 There exists some freedom in the determination of the relation
 between $|T_\mu\ra$ and $|\ov T_\mu\ra$.
 It proves convenient to assume that the basis vectors $|T_\mu\ra $ are ``real'',
 that is,
\begin{equation}\label{ovT=T}
 {|\ov T_\mu\ra }=|T_\mu\ra .
\end{equation}

 Then, the complex conjugates of $|K\ra$ and $\la K|$ are written as
\begin{gather}\label{|ovK>}
 |\ov K\ra = K^{\mu *} |T_\mu\ra,
 \quad \la \ov K | = \la T_\mu| K^{\mu *}.
\end{gather}
 The second scalar product in Eq.(\ref{Kmu-Jmu}) is written as
\begin{gather}\label{<Kmu-Jmu*>}
 \la \ov J |K\ra = J^*_{\mu} K^\mu,
\end{gather}
 and it is easy to verify that
\begin{gather}\label{JK-KJ*}
 \la\ov K|J\ra ^* = \la\ov J|K\ra.
\end{gather}

 It proves convenient to introduce an operator related to the EM-symbols,
 denoted by $\sigma$, namely,
\begin{equation}\label{sigma}
 \sigma := \sigma^{\mu AB'} |T_\mu\ra \la S_{B'A}|.
\end{equation}
 Using $\ov\sigma$ to indicate the complex conjugate of $\sigma$,
 Eq.(\ref{sigma}) gives
\begin{equation}\label{ov-sigma}
 \ov\sigma = \ov\sigma^{\mu A'B} |T_\mu\ra \la S_{BA'}|,
\end{equation}
 where $\ov\sigma^{\mu A'B} = (\sigma^{\mu AB'})^*$.
 Making use of the explicit expressions of the EM-symbols in Eq.(\ref{sigma^AB}),
 it is easy to verify that
\begin{equation}\label{sig-c}
 \ov\sigma_{\mu}^{B'A}= \sigma_{\mu}^{AB'}.
\end{equation}
 One can show that $\ov \sigma = -\sigma $.

\section{Vacuum contribution to the connection field}\label{sect-vacuum-contribution}

 In this section, after recalling some basic contents of the electroweak theory in Sec.\ref{sect-GWS},
 we discuss a possible contribution to the connection field from vacuum fluctuations 
 of the leptonic fields in Sec.\ref{sect-Avs}.
 Finally, in Sec.\ref{sect-gt-Avs} we discuss gauge transformations for this part of the connection field.

\subsection{The Lagrangian and gauge transformations in the electroweak theory}\label{sect-GWS}

 We use a label $\eta $ to indicate species of leptons,
 with $\eta=e$ for electron and $\eta=\nu$ for electron neutrino,
 and use $\ov \eta$ to indicate the antiparticle of $\eta$.
 The label $\eta$ does not obey the convention of repeated label implying a summation over it
 and, when a summation over $\eta$ is performed, we write explicitly $\sum_\eta$.
 In terms of creation and annihilation operators, the quantized fields for leptons are written as
\begin{subequations}\label{psi-psi+-QED}
\begin{gather}\label{psi-QED}
  \psi_\eta(x) = \int d\ww p \left(  b^r_\eta(\bp) U^{r}_\eta(\bp) e^{-ipx}
  + d^{r\dag}_{\ov\eta}(\bp) V^{r}_\eta(\bp) e^{ipx} \right),
\\ \psi^\dag_\eta(x) = \int d\ww p \left(  b^{r\dag}_\eta(\bp)
 U^{\dag r}_\eta(\bp) e^{ipx} +d^r_{\ov\eta}(\bp) V^{\dag r}_\eta(\bp)e^{-ipx} \right), \label{psi+-QED}
\end{gather}
\end{subequations}
 where $d\ww p = \frac{1}{p^0} d^3p$ \cite{footnote-prefactor}.
 Note that $U^{r}_\eta(\bp) $ etc.~are just the Dirac spinors discussed in the previous section,
 but with a subscript $\eta$ indicating the species of particle they are related to.
 The creation and annihilation operators satisfy the well-known anticommutation relations,
\begin{subequations}
\begin{gather}
  \{ b^{\dag r}_\eta(\bp) ,  b^{\dag s}_\eta(\bq ) \} =0,
 \\  \{ d^{\dag r}_\eta(\bp) ,  d^{\dag s}_\eta(\bq ) \} =0,
 \\  \{ b^r_\eta(\bp) ,  b^{\dag s}_\eta(\bq ) \} =p^0 \delta^{rs} \delta^3(\bp-\bq), \label{b-bdag=0}
 \\  \{ d^r_\eta(\bp) ,  d^{\dag s}_\eta(\bq ) \} =p^0 \delta^{rs} \delta^3(\bp-\bq), \label{d-ddag=0}
\end{gather}
\end{subequations}
 and $\{ b^{r\dag}_\eta(\bp) , d^{s\dag}_{\ov\eta}(\bq) \} =0$ for $r,s=0,1$.

 The photonic field is written as
\begin{gather}
 A_\mu(x) = \int d\ww k  a_{\lambda}(\bk) \varepsilon^{(\lambda)}_\mu(\bk) e^{-ikx}
  + a_{\lambda}^{\dag}(\bk) \varepsilon^{(\lambda)*}_\mu(\bk) e^{ikx}, \label{Amu-QED}
\end{gather}
 and other bosonic fields can be written in a similar way.
 As is well known, components of the bosonic fields are not completely independent of each other.
 But, this fact does not influence the discussions to be
 given below and we do not need to give any further discussion for it.

 We use $\psi_{L\eta}(x)$ and $\psi_{R\eta}(x)$ to indicate the LH part
 and the RH part of $\psi_\eta(x)$ in Eq.(\ref{psi-QED}), respectively.
 In a doublet form,  the LH fields are written as
\begin{gather}\label{EL}
 E_L(x) = \left( \begin{array}{c} \psi_{L \nu }(x) \\  \psi_{L e }(x) \end{array} \right).
\end{gather}
 The two-component-spinor expressions of Dirac spinors given in the previous section,
 e.g., that in Eq.(\ref{Up-uv-main}), are quite convenient for discussions to be given below,
 because they are already written in the LH-RH forms.

 The total Lagrangian density includes two parts,
\begin{gather}\label{Lag}
 \LL(x) = \LL_{l}(x) + \LL_{b}(x),
\end{gather}
 where $\LL_{l}(x)$ represents the part involving leptons and $\LL_{b}(x) $ indicates
 the part for bosonic Yang-Mills fields.
 The part $\LL_{b}(x) $ can be constructed from commutators of covariant derivatives.
 The part $\LL_{l}(x)$ includes two subparts, one for the LH fields, denoted by $\LL_l^{\rm LH}(x)$,
 and the other for the RH field, denoted by $\LL_l^{\rm RH}(x)$,
 $\LL_{l}(x) = \LL_l^{\rm LH}(x) + \LL_l^{\rm RH}(x)$, where
\begin{subequations}\label{LL-l-LR}
\begin{gather}\label{LL-LH}
 \LL_l^{\rm LH}(x) = i E_L^\dag \gamma^0 \gamma^\mu D_{\mu} E_L,
 \\ \LL_l^{\rm RH}(x) = i \psi_{Re}^\dag \gamma^0 \gamma^\mu D_{\mu} \psi_{Re}. \label{LL-RH}
\end{gather}
\end{subequations}
 The covariant derivative $D_\mu$ is written as
\begin{gather}\label{Dmu}
 D_\mu = \pp_\mu -i A^{\rm con}_\mu(x),
\end{gather}
 where $A^{\rm con}_\mu(x)$ represents the connection field.
 When $D_\mu$ acts on the LH and RH parts of the leptonic fields,
 the field $A^{\rm con}_\mu(x)$ has different expressions,
 which we indicate as $A^{\rm con}_{L,\mu}(x)$ and $A^{\rm con}_{R,\mu}(x)$, respectively. 
 Specifically, 
\begin{subequations}\label{Acon}
\begin{gather}\label{Dmu-L}
 A^{\rm con}_{L,\mu}(x) = g  A^{a}_\mu(x) \tau^a  + y g'B_\mu(x),
 \\ A^{\rm con}_{R,\mu}(x) = y g'B_\mu(x),
\end{gather}
\end{subequations}
 where $y=1/2$ for $A^{\rm con}_{L,\mu}(x)$ and $y=1$ for $A^{\rm con}_{R,\mu}(x)$,
 $\tau^a$ of $a=1,2,3$ indicate the three generators of the $SU(2)$ group,
 which can be written as $\tau^a_{\eta'\eta}$ in the matrix form,
 $A_\mu^a(x)$ represent  the bosonic fields related to the $SU(2)$ symmetry,
 and $B_\mu(x)$ is for the bosonic field related to $U(1)$.

 Gauge transformations with respect to the $U(1)$ group take the following forms,
\begin{subequations}\label{U1-gt}
\begin{gather}\label{psi-eta-gt}
 \psi_\eta(x) \to  V_1(x) \psi_\eta(x),
 \\ \psi_\eta^\dag(x) \to  \psi_\eta^\dag (x)  V_1^*(x), \label{psi-eta-gt+}
 \\ B_\mu(x) \to B_\mu(x) - \frac{i}{yg'} [\pp_\mu V_1(x)]  V_1^*(x), \label{Bamu-gt}
\end{gather}
\end{subequations}
 where $V_1(x)$  represents $U(1)$ elements,
\begin{gather}\label{Vx}
 V_1(x) = \exp \big( -i \alpha(x) \big)
\end{gather}
 with a $c$-number real function  $\alpha(x)$.
 Under these $U(1)$ transformations, the bosonic fields $A^a_\mu(x)$ do not change
 and the connection field changes as
\begin{gather}\label{Acon-U1}
 A^{\rm con}_\mu(x) \to A^{\rm con}_\mu(x) - i [\pp_\mu V_1(x)]  V_1^*(x),
\end{gather}
 for both the LH and RH parts of the leptonic fields.
 Note that, in the LH case, the second term on the rhs of Eq.(\ref{Acon-U1})
 in fact contains a $2\times 2$ unit matrix for the label $\eta$, which is omitted here and hereafter for brevity.

 Gauge transformations of the $SU(2)$ group take the following forms,
\begin{subequations}\label{SU2-gt}
\begin{gather}\label{EL-gt}
 E_L(x) \to  V_2(x) E_L(x),
 \\ E_L^\dag(x) \to  E_L^\dag (x)  V_2^\dag(x), \label{EL-gt+}
 \\ A^a_\mu(x) \tau^a \to V_2 A^a_\mu(x) \tau^a  V_2^\dag
 - \frac ig (\pp_\mu V_2)  V_2^\dag, \label{Aamu-gt}
\end{gather}
\end{subequations}
 where $V_2(x)$  represents $SU(2)$ elements,
\begin{gather}\label{Vx}
 V_2(x) = \exp \big( -i \alpha^a(x) \tau^a \big)
\end{gather}
 with $c$-number real functions  $\alpha^a(x)$.
 Under these $SU(2)$ transformations, the bosonic field $B_\mu(x)$ does not change,
 and the connection field changes as
\begin{gather}\label{Acon-SU2}
 A^{\rm con}_{L,\mu}(x) \to V_2 A^{\rm con}_{L,\mu}(x)  V_2^\dag
 - i (\pp_\mu V_2)  V_2^\dag.
\end{gather}
 The total Lagrangian is invariant under both classes of gauge transformations discussed above.

\subsection{A vacuum contribution to the total connection}\label{sect-Avs}

 We assume that, besides the bosonic vector fields $A^a_\mu(x)$ and $B_\mu(x)$,
 the total connection that should appear in the covariant derivative in Eq.(\ref{Dmu}) may contain
 a contribution from vacuum fluctuations of the leptonic fields.
 We call this contribution \emph{vacuum vector fields}, and use
 $A^{\rm vs}_{L,\mu}$ and $A^{\rm vs}_{R,\mu}$ to denote them,
 which are related to the LH and RH leptonic fields, respectively, 
 with the superscript ``vs'' standing for ``vacuum state''.
 Specifically, we propose that, instead of Eq.(\ref{Dmu}), the covariant derivative $D_\mu$ should be written as
\begin{gather}\label{Dmu-tot}
 D_\mu = \pp_\mu -i A^{\rm tot-c}_{L(R),\mu}(x),
\end{gather}
 where $A^{\rm tot-c}_{L(R),\mu}(x)$ indicates the total connection,
\begin{gather}\label{Acon-total-F}
 A^{\rm tot-c}_{L(R),\mu}(x) = A^{\rm con}_{L(R),\mu}(x) + A^{\rm vs}_{L(R),\mu}.
\end{gather}

 The field $A^{\rm vs}_{L,\mu}$ should be written as a $2\times 2$ matrix,
 in accordance with the doublet form of $E_L$ for the LH fields [cf.~Eq.(\ref{LL-LH})], that is, 
\begin{gather}\label{}
 A^{\rm vs }_{L, \mu} =  \left( \begin{array}{cc}    A^{\rm vs }_{L\nu \nu, \mu} & A^{\rm vs }_{L\nu e, \mu}
 \\ A^{\rm vs }_{Le \nu, \mu} &   A^{\rm vs }_{L ee, \mu} \end{array} \right).
\end{gather}
 where $A^{\rm vs }_{L \eta \eta', \mu}$ indicates the vacuum vector field
 that is related to the species $\eta$ and $\eta'$.
 Meanwhile, the field $A^{\rm vs}_{R,\mu}$ has a singlet form and, for one species $\eta$, 
 we write it as $A^{\rm vs}_{R\eta,\mu}$.
 Since the vacuum vector fields are due to vacuum fluctuations,
 it is natural to write them as the vacuum expectation values of some
 Lorentz-covariant operators, which we denote by  $\A^{\rm vs }_{\ldots, \mu}$, specifically, 
\begin{subequations}\label{A-opA}
\begin{gather}\label{A-opA-L}
 A^{\rm vs }_{L\eta \eta' , \mu} = \N \la 0|\A^{\rm vs }_{L\eta \eta' , \mu} |0\ra,
 \\ A^{\rm vs }_{R \eta, \mu} = \N \la 0|\A^{\rm vs }_{R\eta, \mu} |0\ra,
\end{gather}
\end{subequations}
 where $|0\ra$ indicates the vacuum state and $\N$ is a constant prefactor.
 For the sake of convenience in presentation, sometimes we simply write $A^{\rm vs }_{L \eta\eta , \mu}$
 as $A^{\rm vs }_{L \eta, \mu}$.

 From the expression of the leptonic fields in Eq.(\ref{psi-QED}) and the expressions of the
 Dirac spinors $U^r(\bp)$ and $V^r(\bp)$ in Eqs.(\ref{Up-uv-main}) and (\ref{Vp-uv-main}),
 one gets the following expression for LH leptonic fields,
\begin{gather}
  \psi_{L \eta }^A(x) = \frac{1}{\sqrt 2} \int d\ww p \ u^{r,A}_{\eta }(\bp) G^{r}_{L\eta}(\bp,x) \label{psiL}
\end{gather}
 with $\eta =\nu, e$, where  $G^{r}_{L\eta}(\bp,x)$ is defined by
\begin{gather}
  G^{r}_{L\eta}(\bp,x) =  b^r_\eta(\bp) e^{-ipx}  + d^{r\dag}_{\ov \eta}(\bp)  e^{ipx}. \label{G-eta}
\end{gather}
 Note that spinor states in the LH fields are two-component spinors and,
 hence, $\psi$ has an explicit spinor label $A$.
 The conjugate fields are written as
\begin{gather}
 \psi_{L \eta }^{\dag A'}(x) = \frac{1}{\sqrt 2}  \int d\ww p \ \ov u^{r,A'}_{\eta}(\bp) \ov G^{r}_{L\eta}(\bp,x), \label{psiL+}
\end{gather}
 where
\begin{gather}
 \ov G^{r}_{L\eta}(\bp,x) = b^{r\dag}_\eta(\bp) e^{ipx}  +d^r_{\ov \eta}(\bp) e^{-ipx} . \label{ov-G-eta}
\end{gather}
 Similarly, the RH fields are written as
\begin{gather}
  \psi_{R\eta,A'}(x) = \frac{1}{\sqrt 2} \int d\ww p \ \ov v^{r}_{\eta,A'}(\bp) G^{r}_{R\eta}(\bp,x), \label{psiRe}
 \\ \psi_{R\eta,A}^{\dag }(x) = \frac{1}{\sqrt 2}  \int d\ww p \ v^{r}_{\eta,A}(\bp) \ov G^{r}_{R\eta}(\bp,x), \label{psiRe+}
\end{gather}
 where
\begin{gather}
  G^{r}_{R\eta}(\bp,x) =  b^r_\eta(\bp) e^{-ipx}  - d^{r\dag}_{\ov \eta}(\bp)  e^{ipx}, \label{GR-eta}
 \\ \ov G^{r}_{R\eta}(\bp,x) = b^{r\dag}_\eta(\bp) e^{ipx}  - d^r_{\ov \eta}(\bp) e^{-ipx} . \label{ov-GR-eta}
\end{gather}

 The vector feature of the fields $A^{\rm vs }_{L\eta, \mu}$ suggests that
 they may include either a term $\pp_\mu \psi_{L\eta}^{A}$ or a term $\pp_\mu \psi_{L\eta}^{\dag A'}$.
 We consider the case of $\pp_\mu \psi_{L\eta}^{A}$ in what follows and similar for
 $A^{\rm vs }_{R\eta, \mu}$.
 In order to construct an operator $\A^{\rm vs }_{L\eta, \mu}$,
 whose vacuum expectation value does not necessarily vanish,
 besides the term $\pp_\mu \psi_{L\eta}^A$, another term is needed that contains 
 creation/annihilation operators like those in $\psi_{L\eta}^{\dag }(x)$.
 We use $\phi_{L\eta}^{\dag }(x)$ to indicate this second term.
 Since the operator $\A^{\rm vs }_{L\eta, \mu}$, as a part of the connection,
 should not depend on the spinor label $A$ in $\pp_\mu \psi_{L\eta}^A$, $\phi_{L\eta}^{\dag }(x)$
 should contain a spinor label of the same type,
 such that a scalar product can be formed [cf.~Eq.(\ref{<chi|kappa>})] \cite{footnote-Avsmu}.

 Specifically, the spinor-label dependence of the fields $\psi_{L\eta}^A$ comes from 
 the label $A$ of the Weyl spinor $u^{r,A}_{\eta }(\bp)$ in them. 
 The simplest way of getting rid of this dependence is to let $\phi_{L\eta}^{\dag }(x)$
 contain a Weyl spinor of the form $w^{s}_{\eta,A}(\bq)$ with the label $A$ in the lower position, 
 such that $\A^{\rm vs }_{L\eta, \mu}$ contains a product term $u^{r,A}_{\eta }(\bp) w^{s}_{\eta,A}(\bq) $.
 In fact, making use of Eqs.(\ref{I})-(\ref{kappa-A}), this term is written as
\begin{gather}\label{wu-abs}
 - \la w^{s}_{\eta}(\bq)|S_A\ra  \la S^A|u^{r}_{\eta }(\bp)\ra
 = \la  w^{s}_{\eta}(\bq)|I_\WW | u^{r}_{\eta }(\bp)\ra,
\end{gather}
 where the superficial dependence on $A$ is moved away as seen in the identity operator $I_\WW$. 
 Therefore, we write the fields $\phi_{L\eta}^{\dag }(x)$ in the following form,
\begin{gather}
  \phi_{L \eta,A}^{\dag }(x) = \frac{1}{\sqrt 2}  \int d\ww p \ w^{r}_{\eta,A}(\bp) \ov G^{r}_{L\eta}(\bp,x).
  \label{phi-Leta+}
\end{gather}
 For a reason that will be discussed in the paragraph below Eq.(\ref{Avs-LH-gt}), we set that
\begin{gather}\label{s-v}
 |w^{r}_{\eta}(\bp) \ra = |v^{r}_{\eta}(\bp) \ra.
\end{gather}

 It is straightforward to generalize the above discussions to the operators $\A^{\rm vs }_{L \eta \eta', \mu}$
 related to two species of lepton, as well as to $\A^{\rm vs }_{R\eta, \mu}$,
 where the latter contain terms 
\begin{gather}
 \phi_{R \eta}^{\dag A'}(x) = \frac{1}{\sqrt 2}  \int d\ww p \ \ov u^{r,A'}_{\eta}(\bp) 
 \ov G^{r}_{R\eta}(\bp,x). \label{phiR+}
\end{gather}
 To summarize, we propose to consider the following explicit expressions for the $\A$-operators,
\begin{subequations}\label{op-Avs}
\begin{gather}\label{}
 \A^{\rm vs }_{L\eta\eta', \mu} =  (\pp_\mu \psi^A_{L\eta}) \phi_{L\eta',A}^{\dag }, \label{op-Avs-L}
 \\  \A^{\rm vs }_{R\eta, \mu} =  (\pp_\mu \psi_{R\eta,A'}) \phi_{R\eta}^{\dag A'}. \label{op-Avs-R}
\end{gather}
\end{subequations}
 Note that, similar to the case with $A^{\rm vs }$ discussed previously,
 $\A^{\rm vs }_{L\eta\eta, \mu} =\A^{\rm vs }_{L\eta, \mu}$.
 For brevity, we call the fields in Eqs.(\ref{phi-Leta+}) and (\ref{phiR+}) \emph{$\phi^\dag$-fields}, 
 while, call fields like that in Eq.(\ref{psiL}) \emph{$\psi$-fields}.

 Substituting the above-discussed plane-wave expansions of  fields into Eq.(\ref{op-Avs})
 and making use of Eq.(\ref{b-bdag=0}),
 as well as Eq.(\ref{ip-vu}) which gives that $v^{r}_{\eta,A}(\bp)   u^{r,A}_\eta (\bp) =2$,
 direct derivation shows that
\begin{gather}\label{}\label{<Avs>=0}
 \la 0|\A^{\rm vs }_{L\eta, \mu} |0\ra = \la 0|\A^{\rm vs }_{R\eta, \mu} |0\ra  = -i  \int d\ww p \ p_\mu.
\end{gather}
 The integration on the rhs of Eq.(\ref{<Avs>=0}) over the whole region of three-momentum 
 gives a divergent result. 
 In order to get finite results, one may employ a momentum regularization scheme,
 in which a finite three-momentum region with $|\bp| < \Lambda$ is considered \cite{Gu13}.
 Within this momentum region, the integration on the rhs of Eq.(\ref{<Avs>=0}), denoted by
 $ J_\mu^\Lambda$, can be easily evaluated,
\begin{gather}\label{}
 J_\mu^\Lambda = \left(\frac 43 \pi \Lambda^3, 0,0,0 \right).
\end{gather}
 Moreover, it is easy to verify that $\la 0|\A^{\rm vs }_{L\eta \eta', \mu} |0\ra=0$ for
 $\eta \ne \eta'$.
 Thus, the vacuum vector fields are written as
\begin{subequations}\label{Avs-Lambda}
\begin{gather}\label{Avs-Lambda-L}
 A^{\rm vs, \Lambda}_{L\eta \eta', \mu} = -i\N J_\mu^\Lambda \delta_{\eta \eta'},
 \\ A^{\rm vs, \Lambda }_{R\eta, \mu} =-i\N J_\mu^\Lambda.
\end{gather}
\end{subequations}

 Since the rhs of Eq.(\ref{Avs-Lambda}) are $x$-independent and $A^{\rm vs, \Lambda}_{L\eta \eta', \mu}$ 
 include a unit matrix $\delta_{\eta \eta'}$,
 the vacuum vector fields should give a zero contribution to the Lagrangian of the connection fields. 
 Substituting Eq.(\ref{Avs-Lambda}) into Eq.(\ref{LL-l-LR}), one finds that 
 their contribution to the interaction Lagrangian,
 denoted by $\LL^{\rm vs, \Lambda}_{\rm int }$, is written as 
\begin{gather}\label{LL-vs}
 \LL^{\rm vs, \Lambda}_{\rm int } = -i\N \left( \psi_{e}^\dag \gamma^0 \gamma^\mu J_\mu^\Lambda \psi_{e}
 + \psi_{L\nu}^\dag \gamma^0 \gamma^\mu J_\mu^\Lambda \psi_{L\nu} \right). 
\end{gather}
 In the computation of scattering matrix, due of the constant feature of $J_\mu^\Lambda$, 
 the contribution of $\LL^{\rm vs, \Lambda}_{\rm int }$ can be dealt with in a way similar to the 
 mass terms for leptons. 
 Since no experimental signature for existence of such a contribution has ever been observed,
 when a renormalization procedure is carried out, one may introduce a counter term that exactly
 cancels this contribution and, then, take the limit of $\Lambda \to \infty$.
 Under this treatment, the vacuum vector fields have no experimentally-observable effect.

\subsection{Gauge transformations of vacuum vector fields}\label{sect-gt-Avs}

 In this section, we discuss changes of the vacuum vector fields $A^{\rm vs }_{L(R), \mu}$
 under gauge transformations of the leptonic fields.
 Let us first discuss the $U(1)$ transformations in Eq.(\ref{psi-eta-gt}).
 We assume that the $\phi^\dag$-fields transform in the same way as the $\psi^\dag$-fields, that is,
\begin{gather}\label{phi-gt}
 \phi_{L(R)\eta,A(A')}^{\dag } \to \phi_{L(R)\eta,A(A')}^{\dag } V_1^*(x).
\end{gather}
 Substituting these transformations into the expressions of the operators $\A$
 in Eqs.(\ref{op-Avs}), and noting that 
\begin{gather}\label{<0psi-phi0>}
 \la 0|\psi^A_{L\eta} \phi_{L\eta',A}^{\dag }|0\ra = \delta_{\eta\eta'} \int d\ww p
\end{gather}
 with the integration understood in the same way as that discussed
 above for Eq.(\ref{<Avs>=0}), straightforward derivation gives the following change of
 the vacuum vector fields, 
\begin{gather}\label{AvsLR-1}
 A^{\rm vs }_{L(R), \mu} \to A^{\rm vs }_{L(R), \mu} + \N  (\pp_\mu V_1) V_1^* \int d\ww p.
\end{gather}
 We take the prefactor $\N$ as
\begin{gather}\label{}
 \N = -i \left( \int d\ww p \right)^{-1},
\end{gather}
 such that the above change of the vacuum vector fields is written as
\begin{gather}\label{Avs-V1-change}
 A^{\rm vs }_{L(R), \mu} \to A^{\rm vs }_{L(R), \mu} -i(\pp_\mu V_1) V_1^* .
\end{gather}
 This change of the vacuum vector fields has the same form as the rhs of Eq.(\ref{Acon-U1}). 

 Next, we discuss the $SU(2)$ transformations in Eq.(\ref{EL-gt}).
 With the leptonic labels written explicitly, one writes
 $\psi_{L \eta }^A(x) \to V_{2,\eta\eta'} \psi_{L \eta' }^A(x)$.
 We also assume that the $\phi^\dag$-fields transform in the same way as the $\psi^\dag$-fields, that is,
\begin{gather}\label{phi-gt-V2}
 \phi_{L(R)\eta,A(A')}^{\dag } \to \sum_{\eta'} \phi_{L(R)\eta',A(A')}^{\dag } V_{2,\eta'\eta}^\dag.
\end{gather}
 Substituting the above transformations into Eq.(\ref{op-Avs-L}), then, into Eq.(\ref{A-opA}),
 one gets that
\begin{gather}\notag
  A^{\rm vs }_{L\eta \eta' , \mu} \to \sum_{\eta_1 \eta_1'} \N \la 0|\left(
  \pp_\mu V_{2,\eta\eta_1} \psi_{L \eta_1 }^A \right)
 \phi_{L\eta_1',A}^{\dag } V_{2,\eta_1'\eta'}^\dag(x)|0\ra.
\end{gather}
 Then, noting that $ V_2 A^{\rm vs }_{L , \mu} V_2^\dag = A^{\rm vs }_{L , \mu} $ 
 as a result of the term $\delta_{\eta\eta'}$ on the rhs of Eqs.(\ref{Avs-Lambda-L}),
 and making use of Eq.(\ref{<0psi-phi0>}), direct derivation gives that
\begin{gather}\label{Avs-LH-gt}
  A^{\rm vs }_{L, \mu} \to A^{\rm vs }_{L , \mu} -i (\pp_\mu V_{2}) V_{2}^\dag.
\end{gather}
 It is seen that the change in Eq.(\ref{Avs-LH-gt}) is equal to the second term on the rhs of Eq.(\ref{Acon-SU2}).

 Now, we can discuss the reason of setting Eq.(\ref{s-v}).
 If otherwise setting $|w^{r}_{\eta}(\bp) \ra = |u^{r}_{\eta}(\bp) \ra$, 
 following arguments similar to those
 discussed above, one finds that both the vacuum vector fields $A^{\rm vs }_{L(R), \mu}$ and their changes under
 $U(1)$ and $SU(2)$ transformations would be proportional to a term $(u^{r}_{\eta,A}(\bp)   u^{r,A}_\eta (\bp))$,
 which definitely vanishes as a result of the property in Eq.(\ref{ck=-kc}).

 To summarize, with the vacuum vector fields taken into account, 
 the total Lagrangian remains invariant under $U(1)$ and $SU(2)$ gauge transformations of the leptonic fields, 
 if instead of Eqs.(\ref{Bamu-gt}) and (\ref{Aamu-gt}),
 one assumes the following transformations of the bosonic fields,
\begin{subequations}\label{SU-gt-v}
\begin{gather}
 B_\mu(x) \to B_\mu(x), \label{Bamu-gt-v}
 \\ A^a_\mu(x) \tau^a \to V_2(x) A^a_\mu(x) \tau^a  V_2^\dag(x). \label{Aamu-gt-v}
\end{gather}
\end{subequations}

\section{Local phase changes of basis states}\label{sect-QED}

 In this section, we discuss local phase transformations of basis states that are employed
 in the description of single-lepton states, as well as related
 changes of the leptonic and connection fields.
 Specifically, we write the basis states and leptonic fields in the abstract notation in Sec.\ref{sect-QED-abs},
 then, discuss the local phase transformations in Sec.\ref{sect-QED-tran}.
 Within this section, there is no need to distinguish between the two species of $\eta=\nu$ and $\eta=e$
 and, hence, we omit the label $\eta$ for brevity.

\subsection{Basis states in the abstraction notation}\label{sect-QED-abs}

 In this section, we discuss single-leptonic basis states,  
 each of which is a direct product of a spatial part $|\bx\ra$ related to a spatial point $\bx$ and 
 a spinor part of the Dirac type.
 The spinor part of the basis, denoted by $|Q_\alpha\ra$ with $\alpha =A, B'$ ($A=0,1$ and $B'=0',1'$), 
 is defined by
\begin{gather}\label{QA}
 |Q_{\alpha=A}\ra := \left( \begin{array}{c} |S^A \ra \\ 0 \end{array} \right),
 \quad  |Q_{\alpha =B'}\ra := \left( \begin{array}{c} 0 \\ |\ov S_{B'}\ra \end{array} \right).
\end{gather}
 The bras of $|Q_\alpha\ra$, denoted by $\la Q_\alpha|$, are written as
\begin{gather}\label{<Qa|}
 \la Q_A| = (\la S^A |, 0), \qquad   \la Q_{B'}| = ( 0, \la \ov S_{B'}|).
\end{gather}
 Here, no complex conjugation is involved when kets are changed to bras,
 like the case of Weyl spinors [cf.~Eq.(\ref{<kappa|})].

 Components of the Dirac spinors $|U^r(\bp)\ra$ and $|V^r(\bp)\ra$ in the basis $|Q_\alpha\ra$,
 denoted by $U^r_\alpha(\bp)$ and $V^r_\alpha(\bp)$, respectively, are  given by
\begin{subequations}\label{UVr-alpha}
\begin{gather}\label{Ur-alpha}
 U^r_\alpha(\bp) = \la Q_\alpha| U^r(\bp)\ra,
 \\ V^r_\alpha(\bp) = \la Q_\alpha| V^r(\bp)\ra.
\end{gather}
\end{subequations}
 Making use of  Eqs.(\ref{kappa-A}), (\ref{|U-Vp>}), and (\ref{<Qa|}), 
 it is easy to check that
\begin{subequations}\label{UV-alpha}
\begin{gather}\label{}
 U^r_{\alpha = A}(\bp) = \frac{1}{\sqrt 2}  u^{r A}(\bp),
 \ U^r_{\alpha =B'}(\bp) = \frac{1}{\sqrt 2} \ov v^r_{B'}(\bp),
 \\ V^r_{\alpha = A}(\bp) = \frac{1}{\sqrt 2} u^{r A}(\bp),
  V^r_{\alpha = B'}(\bp) = -\frac{1}{\sqrt 2} \ov v^r_{B'}(\bp).
\end{gather}
\end{subequations}
 Thus, $U^r_\alpha(\bp)$ and $V^r_\alpha(\bp)$ are equal to the
 components of $|U^r(\bp)\ra$ and $|V^r(\bp)\ra$ given in Eqs.(\ref{Up-uv-main}) and (\ref{Vp-uv-main}).

 One may  consider the complex conjugates of $|Q_\alpha\ra$, denoted by $|\ov Q_{\alpha'}\ra$
 with $\alpha' = A',B$, which are written as
\begin{subequations}
\begin{gather}\label{|ov-Qa>}
 |\ov Q_{A'}\ra = \left( \begin{array}{c} |\ov S^{A'} \ra \\ 0 \end{array} \right),
 \quad  |\ov Q_{B}\ra = \left( \begin{array}{c} 0 \\ |S_{B}\ra \end{array} \right).
\end{gather}
\end{subequations}
 The spinors $|\ov Q_{\alpha'}\ra$ give a basis, on which $|\ov U^r(\bp)\ra$
 can be expanded (similar for $|\ov V^r(\bp)\ra$).
 Their bras, denoted by $\la \ov Q_{\alpha'}|$, are written as
\begin{gather}\label{<ov-Qa|}
 \la \ov Q_{A'}| = (\la \ov S^{A'} |, 0), \quad   \la \ov Q_{B}| = ( 0, \la S_{B}|).
\end{gather}
 The complex conjugates of the components in Eq.(\ref{UVr-alpha}) are written as
\begin{subequations}\label{ov-UVr-alpha}
\begin{gather}\label{ov-Ur-alpha}
 \ov U^r_{\alpha'}(\bp) = \la \ov Q_{\alpha'}|\ov U^r(\bp)\ra. 
 \\ \ov V^r_{\alpha'}(\bp) = \la \ov Q_{\alpha'}|\ov V^r(\bp)\ra. 
\end{gather}
\end{subequations}

 Making use of the spinors $|Q_\alpha\ra$, basis states for a single lepton are written as
\begin{gather}\label{|b-x-al>}
 |b_\alpha(\bx)\ra :=  |\bx\ra |Q_\alpha\ra, 
\end{gather}
 each with a definite spatial coordinate $\bx$ and a definite spinor label $\alpha$.
 The bra corresponding to $|b_\alpha(\bx)\ra$ is written as
\begin{gather}\label{}
 \la b_\alpha(\bx)\ra| = \la\bx| \la Q_{\alpha}|.
\end{gather}
 It proves also useful to consider the following vectors:
\begin{gather}\label{|b-x-al'>}
  |b_{\ov\alpha'}(\bx)\ra := |\bx\ra |\ov Q_{\alpha'}\ra \ \ 
 \& \ \ \la b_{\ov\alpha'}(\bx)| = \la\bx| \la \ov Q_{\alpha'}|.
\end{gather}


 Finally, we write the leptonic fields in terms of the basis states discussed above.
 To this end, we make use of Eqs.(\ref{UVr-alpha}) and (\ref{ov-UVr-alpha}), and write the plane waves as 
 $e^{i\bp\cdot \bx} = (2\pi)^{3/2} \la \bx |\bp\ra$ and $e^{-i\bp\cdot \bx} = (2\pi)^{3/2} \la \bp |\bx\ra$.
 Noting Eq.(\ref{ck=-kc}), the leptonic fields in Eq.(\ref{psi-psi+-QED}) can be 
 written in the following forms, with the label $\alpha$ written explicitly, 
\begin{subequations}\label{psi-psi+-QED-br}
\begin{gather}\notag
  \psi_\alpha(x) = (2\pi)^{\frac 32} \la b_\alpha(\bx) | \left(  \int d\ww p 
  |U^{r}(\bp)\ra|\bp\ra   e^{-ip^0t} b^r(\bp)  \right)  
 \\ - (2\pi)^{\frac 32} \left( \int d\ww p  e^{ip^0t} d^{r\dag}(\bp) 
 \la V^{r}(\bp)| \la \bp|\right) | b_\alpha(\bx)\ra , \label{psi-QED-br}
\\ \psi^\dag_{\alpha'}(x) = - (2\pi)^{\frac 32} \left(   \int d\ww p   e^{ip^0t}b^{r\dag}(\bp) 
  \la \ov U^{r}(\bp)| \la \bp|\right) |b_{\ov\alpha'}(\bx)\ra  \notag
 \\  + (2\pi)^{\frac 32} \la b_{\ov\alpha'}(\bx)| \left( \int d\ww p
 |\ov V^{ r}(\bp)\ra |\bp\ra e^{-ip^0t} d^r(\bp)  \right). \label{psi+-QED-br}
\end{gather}
\end{subequations}

\subsection{Local phase transformations for basis states}\label{sect-QED-tran}

 In this section, we discuss phase transformations for the basis states discussed above.
 Due to the local feature of the basis states such as $|b_\alpha(\bx)\ra$, phases of the transformations
 may be $x$-dependent. 
 We are interested in whether such transformations of basis states may make
 the leptonic and connection fields change in manners like $U(1)$ gauge transformations.

 Using $\theta(x)$ to indicate local phases apart from the sign, 
 the above-discussed phase transformations introduce factors $e^{\pm i\theta(x)}$ to the basis states. 
 On the rhs of Eq.(\ref{psi-QED-br}), the basis states appear as bras $\la b_\alpha(\bx) |$ in the first part, 
 while they appear as kets $|b_\alpha(\bx)\ra$ in the second part. 
 Due to this property, if one wants $\psi_\alpha(x)$ to change in a way 
 like a $U(1)$ gauge transformation, under which the two parts transform in a same way, 
 the sign before $\theta(x)$ should not be determined by the  spatial part of a basis state,
 but should be determined by the spinor part.
 The situation is similar with $\psi^\dag_{\alpha'}(x)$ in Eq.(\ref{psi+-QED-br}).
 Specifically, a same sign is related to $|Q_\alpha\ra$ and $\la Q_\alpha|$,
 while, the opposite sign is related to  $|\ov Q_{\alpha'}\ra$ and $\la \ov Q_{\alpha'}|$.
 Thus, the phase transformations of the basis states are written as
\begin{subequations}\label{basis-s-trans}
\begin{gather} \label{bx-bra-change} 
 |b_\alpha(\bx) \ra \to e^{-i\theta(x)} |b_\alpha(\bx) \ra,
 \\ \la b_\alpha(\bx) | \to e^{-i\theta(x)} \la b_\alpha(\bx) |, 
 \\ |b_{\ov \alpha'}(\bx)\ra \to e^{i\theta(x)}  |b_{\ov \alpha'}(\bx)\ra,
 \\ \la b_{\ov \alpha'}(\bx)| \to e^{i\theta(x)}  \la b_{\ov \alpha'}(\bx)|.
\end{gather}
\end{subequations}

 The transformations in Eq.(\ref{basis-s-trans}) can, in fact, be written in 
 a more concise form.
 To this end, let us discuss consequences of the following map of the basis spinors $|S^A\ra$,
\begin{gather}\label{SA-to-qA}
 |S^A\ra \to |q^A\ra = e^{-i\theta} |S^A\ra,
\end{gather}
 where $\theta$ is a real parameter. 
 The relation between Eq.(\ref{|kappa>}) and Eq.(\ref{<kappa|}) implies that
 the bras $\la S^A|$ should change as 
\begin{gather}\label{SA-to-qA-bra}
 \la S^A| \to \la q^A| = e^{-i\theta} \la S^A|.
\end{gather}
 Clearly, the spinors $|q^A\ra$ can also be employed as a basis in the Weyl-spinor space $\WW$.
 But, unlike Eq.(\ref{SA-SB}), they satisfies $\la q^A|q^B\ra = e^{-2i\theta} \epsilon^{AB}$.
 In order to keep the expression of scalar product in Eq.(\ref{<chi|kappa>})
 and the expression of identity operator in Eq.(\ref{I}), one may consider the
 following $\varepsilon$-matrices,
\begin{gather}\label{varepsilon}
 \varepsilon^{AB} = e^{-2i\theta} \epsilon^{AB},
 \qquad \varepsilon_{AB} = e^{2i\theta} \epsilon_{AB},
\end{gather}
 for which $\la q^A|q^B\ra = \varepsilon^{AB}$.

 One can use the $\varepsilon$-matrices to raise and lower the spinor labels of the basis spinors $|q\ra$,
 in a way similar to the $\epsilon$-matrices for $|S\ra$ discussed previously. 
 For example, $|q_A\ra = |q^B\ra \varepsilon_{BA}$.
 This implies that
\begin{gather}\label{q-A-low-ch}
 |S_A\ra \to |q_A\ra = e^{i\theta} |S_A\ra, \
 \la S_A| \to \la q_A| = e^{i\theta} \la S_A|,
\end{gather}
 and $\la q_{A}|q_{B}\ra = \varepsilon_{A B}$.
 Taking complex conjugation for Eqs.(\ref{SA-to-qA})-(\ref{SA-to-qA-bra}) and Eq.(\ref{q-A-low-ch}), 
 one gets that
\begin{subequations}
\begin{gather}\label{SA-to-qA-cmp}
 |\ov S^{A'}\ra \to |\ov q^{A'}\ra = e^{i\theta} |\ov S^{A'}\ra,
 \\ \label{ov-q-A-low-ch}  |\ov S_{A'}\ra \to |\ov q_{A'}\ra = e^{-i\theta} |\ov S_{A'}\ra,
\end{gather}
\end{subequations}
 and similar for bras. 
 Making use of the above results,  it is easy to verify that the $Q$-spinors should change as follows,
\begin{subequations}\label{Qa-ch}
\begin{gather}
 |Q_\alpha\ra \to e^{-i\theta} |Q_\alpha \ra,
 \qquad \la Q_\alpha| \to e^{-i\theta} \la Q_\alpha |,
 \\ |\ov Q_{\alpha'}\ra \to e^{i\theta} |\ov Q_{\alpha'} \ra,
 \qquad \la \ov Q_{\alpha'}| \to e^{i\theta} \la\ov Q_{\alpha'}|.
\end{gather}
\end{subequations}

 Now, we back to the phase transformations of basis states in Eq.(\ref{basis-s-trans}).
 Since the map in Eq.(\ref{SA-to-qA}) implies the changes of the $Q$-spinors in Eq.(\ref{Qa-ch}),
 Eq.(\ref{basis-s-trans}) can be written in the following concise form,
\begin{gather}\label{phase-tran-final}
 |S^A\ra \to  e^{-i\theta (x)} |S^A\ra \quad \text{for $|S^A\ra$ in $|b_\alpha(\bx) \ra$.}
\end{gather}
 That is, the phase changes can be effectively assigned to the basis spinors $|S^A\ra$
 in basis states $|b_\alpha(\bx) \ra$.

 It is straightforward to verify that,
 under the transformations in Eq.(\ref{basis-s-trans}), the leptonic fields
 in Eq.(\ref{psi-psi+-QED-br}) change in the following way, 
\begin{subequations}\label{psi-change}
\begin{gather}
  \psi_\alpha(x) \to  e^{-i\theta(x)} \psi_\alpha(x),
 \\ \psi^\dag_\alpha(x) \to  e^{i\theta(x)} \psi^\dag_\alpha(x),
\end{gather}
\end{subequations}
 which have the same forms as the $U(1)$ gauge transformations.
 To study changes of the $\phi^\dag$-fields in Eqs.(\ref{phi-Leta+}) and (\ref{phiR+}),
 let us put these LH and RH parts together and write them in a unified form like
 $\psi^\dag(x)$ in Eq.(\ref{psi+-QED}), i.e.,  
\begin{gather}
  \phi^{\dag }(x)  = \int d\ww p \left(  b^{r\dag}(\bp)
 W^{\dag r}(\bp) e^{ipx} +d^r(\bp) X^{\dag r}(\bp)e^{-ipx} \right), \label{phi+}
\end{gather}
 where 
\begin{gather} \label{W}
 W^r(\bp) = \frac{1}{\sqrt 2} \left( \begin{array}{c} \ov v^{r}_{A'}(\bp) \\ u^{r,B}(\bp) \end{array} \right),
 \ X^r(\bp) = \frac{1}{\sqrt 2} \left( \begin{array}{c} \ov v^{r}_{A'}(\bp) \\ -u^{r,B}(\bp) \end{array} \right).
\end{gather}
 In order to write the spinors $W^r(\bp)$ and $X^r(\bp)$ in the abstract notation, 
 like the basis spinors $|Q_\alpha\ra$
 in Eq.(\ref{QA}), we introduce basis spinors with the spinor labels in the opposite positions,  i.e.,
\begin{gather}\label{QA-up}
 |Q^{\alpha=A}\ra := \left( \begin{array}{c} |S_A \ra \\ 0 \end{array} \right),
 \quad  |Q^{\alpha =B'}\ra := \left( \begin{array}{c} 0 \\ |\ov S^{B'}\ra \end{array} \right),
\end{gather}
 as well as the corresponding bras $\la Q^\alpha|$ and their complex conjugates $|\ov Q^{ \alpha'}\ra$
 and $\la \ov Q^{ \alpha'}|$. 
 Then, the components of $W^{r}(\bp)$ and $X^r(\bp)$ 
 have the following expressions, with the label $\alpha$ written explicitly, 
\begin{gather}\label{Wr-Qa}
 W^{r,\alpha}(\bp) = \la \ov Q^{\alpha'}|W^r(\bp)\ra,
 \ X^{r,\alpha}(\bp) = \la \ov Q^{\alpha'}|X^r(\bp)\ra,
\end{gather}
 where
\begin{gather}\label{}
 |W^r(\bp)\ra = \frac{1}{\sqrt 2} \left( \begin{array}{c} |\ov v^{r}(\bp)\ra \\ | u^{r}(\bp)\ra \end{array} \right),
 \ |X^r(\bp)\ra = \frac{1}{\sqrt 2} \left( \begin{array}{c} |\ov v^{r}(\bp)\ra \\ -| u^{r}(\bp)\ra \end{array} \right).
\end{gather}

 Using the above notations,  the $\phi^\dag$-fields in Eq.(\ref{phi+}) 
 are written in the following form [cf.~Eq.(\ref{psi+-QED-br})], 
\begin{gather}\label{}
 \phi^{\dag, \alpha }(x) = - (2\pi)^{\frac 32} \left(   \int d\ww p   e^{ip^0t}b^{r\dag}(\bp) 
  \la \ov W^{r}(\bp)| \la \bp|\right) |\bx\ra |Q^\alpha\ra \notag
 \\  + (2\pi)^{\frac 32} \la \bx| \la Q^\alpha| \left( \int d\ww p
 |\ov X^{ r}(\bp)\ra |\bp\ra e^{-ip^0t} d^r(\bp)  \right). \label{phi+-abs}
\end{gather}
 Following a procedure similar to that leading to Eq.(\ref{Qa-ch}), one finds that,
 under the transformations in Eq.(\ref{phase-tran-final}), the basis states $|\bx\ra |Q^\alpha\ra$
 and $ \la \bx| \la Q^\alpha|$ should change as follows,
\begin{gather}\label{Qa-up-ch}
 |\bx\ra |Q^\alpha\ra \to e^{i\theta(x)} |\bx\ra|Q^\alpha \ra,
 \ \la \bx| \la Q^\alpha| \to e^{i\theta(x)} \la \bx| \la Q^\alpha |.
\end{gather}
 Then, the $\phi^\dag$-fields in Eq.(\ref{phi+-abs}) should change as
\begin{gather}\label{phi-bs-change}
 \phi^{\dag,\alpha }(x) \to e^{i\theta(x)} \phi^{\dag,\alpha}(x),
\end{gather}
 in a way similar to $\psi^\dag_\alpha(x)$.

 Clearly, the phase transformations in Eq.(\ref{phase-tran-final}) do not influence 
 the bosonic fields $A^a_\mu$ and $B_\mu$.
 But, they do influence the vacuum vector fields $A^{\rm vs}_{L,\mu}$.
 In fact, making use of Eqs.(\ref{psi-change}) and (\ref{phi-bs-change}),
 and following arguments similar to those leading to Eq.(\ref{Avs-V1-change}), one finds that
\begin{gather}\label{Ars-sc}
 A^{\rm vs}_{L(R),\mu}  \to  A^{\rm vs}_{L(R),\mu}   - \pp_\mu \theta,
\end{gather}
 which has the same form as the change in Eq.(\ref{Avs-V1-change}) under $U(1)$ gauge transformations. 
 It is straightforward to verify that the total Lagrangian is invariant under the above changes
 of the basis states and fields.

 \section{Local changes of LH spinor spaces }\label{sect-SU2}

 In this section, we discuss a type of transformations of the spinor
 spaces for LH leptonic states.
 Specifically, the transformations are introduced in Sec.\ref{sect-rotate-LH-bases}.
 Influences of the transformations on the LH leptonic fields
 are studied in Sec.\ref{sect-LH-lepton}, and those on the bosonic fields are discussed in Sec.\ref{sect-LH-bosonic-F}.
 Finally, in Sec.\ref{sect-LH-Avs}, we discuss changes of the vacuum vector fields. 

\subsection{LH Spinor-space transformations for leptons}\label{sect-rotate-LH-bases}

 We use $\WW_\eta$ to denote the space that is spanned by the LH spinor states of a species $\eta$.
 The basis in it is written as $|S^A_\eta\ra$ of $A=0,1$.
 The direct sum of $ \WW_\nu$ and $ \WW_e$, denoted by $\WW_{e\nu }$,
 $\WW_{e\nu } = \WW_\nu \oplus \WW_e$,
 is spanned by the four spinors $|S^{A}_{\eta}\ra$.
 The complex conjugate space of $\WW_\eta$ is written as $\ov\WW_{\eta}$, 
 with a basis $|\ov S^{A'}_{\eta}\ra$.
 Components of spinors in these spaces are written as, for example, 
\begin{subequations}\label{urA-eta-2}
\begin{gather}\label{urA-eta}
 u^{r,A}_{\eta}(\bp) = \la S^A_\eta|u^{r}_{\eta}(\bp)\ra,
 \\ \ov u^{s,B'}_{\eta}(\bq) = \la  \ov S^{B'}_\eta | \ov u^{s}_{\eta}(\bq) \ra. \label{ov-urA-eta}
\end{gather}
\end{subequations}
 The space dual to $\WW_{e\nu } $ is spanned by bras $\la S^{A}_{\eta}|$.
 Generalizing the scalar product in Eq.(\ref{SA-SB})
 and noting that the two subspaces $\WW_e$ and $\WW_\nu$ are for different particles,
 we assume that the above basis spinors have the following scalar products, 
\begin{gather}\label{SA-SB-eta}
  \la S^{A}_{\eta}|S^{B}_{\eta'}\ra = \epsilon^{A B} \delta_{\eta \eta'}.
\end{gather}

 Suppose that $\WW_{e\nu }$ has another direct-sum division,
 written as $\WW_{e\nu } = \ww\WW_1  \oplus \ww\WW_2$,
 where $\ww\WW_\xi$ of  $\xi=1,2$ are two-dimensional spaces,
 each spanned by $|\ww S^A_{\xi}\ra $ of $A=0,1$.
 The spinors $|\ww S^A_{\xi}\ra $ are mixtures of $|S^A_\eta\ra$ for each label $A$,
 satisfying the following relation,
\begin{gather}\label{|SAxi>}
 |S^A_\eta\ra = \sum_{\xi} R^*_{\eta\xi}|\ww S^A_{\xi}\ra = \sum_{\xi} |\ww S^A_{\xi}\ra R^\dag_{\xi\eta},
\end{gather}
 where $R_{\eta\xi}$ represents a $2\times 2$ matrix element of the $SU(2)$ group.
 It is easy to see that the two  subspaces $\ww\WW_1$ and $\ww\WW_2$
 transform in the same way under Lorentz transformations.
 Moreover, one may require that scalar products of the spinors $|\ww S^A_{\xi}\ra $ have the same
 form as those of $|S^A_{\eta}\ra $ given in Eq.(\ref{SA-SB-eta}), that is,
\begin{gather}\label{SA-SB-xi}
  \la \ww S^{A}_{\xi}|\ww S^{B}_{\xi'}\ra = \epsilon^{A B} \delta_{\xi \xi'}.
\end{gather}

 The relation between the bras $\la \ww S^A_{\xi}|$ and the bras $\la S^A_\eta|$ for a fixed label $A$,
 subject to the relation in Eq.(\ref{|SAxi>}), needs not be like that in Eq.(\ref{<kappa|}),
 because the two bras $\la S^A_\eta|$ of $\eta=\nu, e$ do not span a space dual to $\WW$.
 Generically, one may write
\begin{gather}\label{<SAxi|-1}
 \la S^A_\eta| = \sum_{\xi} X_{\eta\xi}\la \ww S^A_{\xi}|.
\end{gather}
 Substituting Eqs.(\ref{|SAxi>}) and (\ref{<SAxi|-1}) into Eq.(\ref{SA-SB-eta}) and making use of
 Eq.(\ref{SA-SB-xi}), direct derivation gives that $\sum_{\xi}
 R^*_{\eta'\xi}X_{\eta\xi}  = \delta_{\eta \eta'}$.
 This implies that $X_{\eta\xi} = (R^{-1}_{\xi\eta})^* =  R_{\eta \xi}$ and ,  as a result, 
\begin{gather}\label{<SAxi|}
 \la S^A_\eta|   = \sum_{\xi} R_{\eta \xi} \la \ww S^A_{\xi}|.
\end{gather}

 The two subspaces $\ww\WW_1$ and $\ww\WW_2$ discussed above possess all those mathematical properties of
 $\WW_e$ and $\WW_\nu$ that are necessary for the purpose of describing LH spinor states of leptons. 
 Hence, it should be equally legitimate to employ $\ww\WW_1$ and $\ww\WW_2$ 
 to describe LH spinor states of leptons.
 In other words, one may perform a transformation of the description spaces for LH spinor states of leptons
 from $\WW_\eta$ to $\ww\WW_{\xi}$, namely,
\begin{gather}\label{WWeta-WWxi}
 \WW_\eta \to \ww\WW_\xi .
\end{gather}
 We call such a transformation \emph{an LH spinor-space transformation}.
 Hereafter, for the sake of convenience in discussion, instead of $\xi=1,2$, we write $\xi = \nu, e$,
 and we use tilde to indicate results of such a transformation, except for $d\ww p = d^3p/p^0$.

 We are to show that if the LH spinor-space transformations are required to satisfy the following rules,
 then,  they lead to results with formal similarity to $SU(2)$ gauge transformations.
\begin{itemize}
  \item LH spinor-space transformation (LST) rules:
  \\ (i) Basis spinors should change according to Eqs.(\ref{|SAxi>}) and (\ref{<SAxi|});
 \\ (ii) basis-independent terms should be transformed in a directly way,
 with $\eta$ directly replaced by $\xi$;
  \\ (iii) spinors in the $\psi$- and $\psi^\dag$-fields should be written in the form of Eq.(\ref{urA-eta-2}).
\end{itemize}
 Some explanations. 
 (a) As an example of the LST rule-(ii), a spinor $|u^{r}_{\eta}(\bp)\ra $ transforms as 
\begin{gather}\label{u-to-wwu}
 |u^{r}_{\eta}(\bp)\ra \to |\ww u^{r}_{\xi}(\bp)\ra,
\end{gather}
 satisfying
\begin{gather}\label{Su-wwSu}
 \la \ww S^A_{\xi}|\ww u^{r}_{\xi}(\bp)\ra = \la S^A_{\eta}|u^{r}_{\eta}(\bp)\ra
 \quad \text{for $\xi = \eta$},
\end{gather}
 that is, $\ww u^{r,A}_{\xi}(\bp) = u^{r,A}_{\eta}(\bp)$ in the component form.
 The transformed annihilation and creation operators, say, $\ww b^r_{\xi}(\bp)$
 and $\ww b^{\dag r}_{\xi}(\bp)$,
 should satisfy the same anticommutation relations as the old ones.
 (b)  Since $u^A =\la S^A|u\ra = -\la u|S^A\ra$, when writing the spinor components 
 in the LH leptonic fields as scalar products, both $|S^A\ra$ and $\la S^A|$ may be used,
 however, $|S^A\ra$ and $\la S^A|$ transform differently according to Eq.(\ref{|SAxi>}) 
 and Eq.(\ref{<SAxi|}), respectively.
 The LST rule-(iii) requires that spinor components in the leptonic fields should be written in the way 
 that basis spinors appear as bras.

 Finally, the total state space for LH states of single leptons, denoted by $\E_{\rm LH}$, 
 is a direct-product space written as
\begin{gather}\label{ss-LH-lepton}
 \E_{\rm LH} = \bigoplus_{\bx} |\bx\ra \otimes \WW_{e\nu}
 = \bigoplus_{\bx,\eta, A} |\bx\ra |S^A_\eta\ra .
\end{gather}
 When an LH spinor-space transformation is performed, it is in fact done within a
 subspace of $\E_{\rm LH}$, namely, within $|\bx\ra \otimes \WW_{e\nu}$.
 Hence, the transformation may be different at different spatial-temporal points $x$, 
 that is, the matrix $R_{\eta\xi}$ in Eq.(\ref{|SAxi>}) may be $x$-dependent and be written as $R_{\eta\xi}(x)$.
 But, for the sake of simplicity in presentation, in what follows we do not 
 indicate this $x$-dependence explicitly, and we do not write 
 the spatial basis states $|\bx\ra$ explicitly, either.

\subsection{Changes of LH leptonic fields}\label{sect-LH-lepton}

 In this section, we discuss changes of the LH leptonic fields under LH spinor-space transformations.
 Let us first discuss the field $\psi_{L \eta }^A(x) $ in Eq.(\ref{psiL}).
 With the Weyl spinors written as in Eq.(\ref{urA-eta}), 
 according to the LST rules-(i) and (ii), $u^{r,A}_{\eta}(\bp)$ should change to the following summation, 
\begin{gather}\label{}
 \sum_{\xi} R_{\eta \xi} \la \ww S^A_{\xi}|\ww u^{r}_{\xi}(\bp)\ra
 = \sum_{\xi} R_{\eta \xi} \ww u^{r,A}_{\xi}(\bp).
\end{gather}
 As a result, the fields $\psi_{L\eta}^A$ change as
\begin{gather}\label{psiL-change}
 \psi_{L \eta }^A(x) \to  \sum_{\xi} R_{\eta\xi}  \ww \psi_{L \xi }^A(x),
\end{gather}
 where
\begin{gather}
  \ww \psi_{L \xi }^A(x) =  \frac{1}{\sqrt 2} \int d\ww p \ \ww u^{r,A}_{\xi }(\bp) \ww G^{r}_{L\xi}(\bp,x), \label{psiL-xi}
 \\ \ww G^{r}_{L\xi}(\bp,x) = \ww b^r_\xi(\bp) e^{-ipx}  + \ww d^{r\dag}_{\ov \xi}(\bp)  e^{ipx}.
\end{gather}

 Next, we discuss the conjugate fields $\psi_{L\eta}^{\dag B'}(x)$.
 Writing the components $\ov u^{s,B'}_{\eta}(\bq)$ as in Eq.(\ref{ov-urA-eta})
 and noting that the complex conjugate of Eq.(\ref{<SAxi|}) gives that
\begin{gather}\label{<ov-SAxi|}
 \la \ov S^{A'}_\eta| = \sum_{\xi} \la \ww {\ov S}^{A'}_{\xi}|  R^\dag_{\xi \eta},
\end{gather}
 one finds that
\begin{gather}\label{psiL+-change}
 \psi_{L \eta }^{\dag B'}(x) \to  \sum_{\xi}  \ww \psi_{L \xi }^{\dag B'}(x) R_{\xi\eta}^\dag,
\end{gather}
 where
\begin{gather}
 \ww \psi_{L \xi }^{\dag B'}(x) = \frac{1}{\sqrt 2}  \int d\ww p \ \ww{\ov u}^{r,B'}_{\xi}(\bp)
 \ww { \ov G}^{r}_{L\xi}(\bp,x), \label{psiL+-xi}
 \\ \ww {\ov G}^{r}_{L\xi}(\bp,x) = \ww b^{r\dag}_\xi(\bp) e^{ipx}  + \ww d^{r}_{\ov \xi}(\bp)  e^{-ipx}.
\end{gather}
 It is seen that the changes of the LH leptonic fields in Eqs.(\ref{psiL-change}) and (\ref{psiL+-change})
 have the same formal forms as the $SU(2)$ gauge transformations in Eqs.(\ref{EL-gt}) and (\ref{EL-gt+}).

 Finally, we discuss the $\phi^\dag$-fields $\phi_{L \eta,A}^{\dag }(x)$.
 The sole function of these fields is that they are used in the construction of the vacuum vector fields
 [cf.~Eq.(\ref{op-Avs})].
 Hence,  the spinors $w^{r}_{\eta,A}(\bp)$ in them should be written in a form
 that is consistent with the left-hand side of Eq.(\ref{wu-abs}).
 This implies that one should write
\begin{gather}\label{v-vr-SA}
  w^{r}_{\eta,A}(\bp) = -\la w^{r}_{\eta}(\bp)|S_{\eta,A}\ra.
\end{gather}
  Then, making use of Eq.(\ref{|SAxi>}) with the spinor label lowered and Eq.(\ref{s-v}), one gets that
\begin{gather}\label{phiL+-change}
 \phi_{L \eta,A}^{\dag }(x) \to  \sum_{\xi}  \ww \phi_{L \xi,A}^{\dag }(x) R_{\xi\eta}^\dag,
\end{gather}
 where
\begin{gather}
 \ww \phi_{L \xi,A}^{\dag }(x) = \frac{1}{\sqrt 2}  \int d\ww p \ \ww{v}^{r,A}_{\xi}(\bp)
 \ww { \ov G}^{r}_{L\xi}(\bp,x). \label{phiL+-xi}
\end{gather}

\subsection{Changes of bosonic fields}\label{sect-LH-bosonic-F}

 There is no reason for the two bosonic fields $A^a_\mu(x)$ and $B_\mu(x)$ to change
 under LH spinor-space transformations and, hence, they should remain invariant.
 But, because of $\tau^a$, the term $A^a_\mu(x) \tau^a$, which appears in the connection field,  should change.
 In this section, we discuss changes of $A^a_\mu(x) \tau^a$.

 In order to determine the way in which $A^a_\mu(x) \tau^a$ should change under
 the LST rules, we consider the following quantity,
\begin{gather}\label{F-eta-eta'}
 F_{\eta' \eta,B'A} = \ov\sigma^{\mu}_{B'A}  A_\mu^a \tau^a_{\eta'\eta}.
\end{gather}
 To write $F_{\eta' \eta,B'A}$ in the abstract notation, let us consider 
 the three terms in it separately.
 Related to $\ov\sigma^{\mu}_{B'A}$ is the operator $\ov\sigma$ in Eq.(\ref{ov-sigma}), 
 which is written as follows for a species $\eta$,
\begin{gather}\label{sigma-eta}
 \ov\sigma_\eta = |T_\mu\ra \ov\sigma^{\mu}_{ D'C} \la S^{C}_{\eta}| \la \ov S^{D'}_{\eta}|,
\end{gather}
 where Eq.(\ref{f-AB}) has been used to move positions of the spinor labels.
 Related to the matrix $\tau^a_{\eta'\eta}$, one may consider the following operator,
\begin{gather}\label{pi-a}
 \pi^a := \sum_{\eta,\eta'} \tau^a_{\eta'\eta} |S^A_{\eta'}\ra \la S_{\eta,A}|.
\end{gather}
 The bosonic fields $A^a_\mu(x)$ can be written in a form like that in Eq.(\ref{Amu-QED}),
 which contain the polarization vectors $\varepsilon^{\lambda}_\mu(\bk)$.
 These vectors are written as $|\varepsilon^{\lambda}(\bk)\ra$ in the abstract notation,
 with $|\varepsilon^\lambda(\bk)\ra = \varepsilon^{\lambda}_\mu(\bk) |T^\mu\ra$.
 Making use of Eqs.(\ref{K-mu})-(\ref{<K|J>=<J|K>}) and (\ref{ovT=T}), direct derivation shows that
\begin{gather}\label{eps-<>}
 \varepsilon^{\lambda}_\mu(\bk) = \la \varepsilon^\lambda(\bk) |  T_\mu \ra, \quad
 \varepsilon^{\lambda*}_\mu(\bk) = \la \ov \varepsilon^{\lambda}(\bk)|T_\mu\ra.
\end{gather}
 Hence, the fields $A^a_\mu(x)$ can be written in the following form,
\begin{gather}\label{Aa-Tmu}
 A^a_\mu \equiv A^a_{\la \VV |}|T_\mu\ra,
\end{gather}
 where $A^a_{\la \VV |}$  include bra-vectors that can form scalar products with $|T_\mu\ra$.

 Making use of the operators discussed above, $F_{\eta' \eta,B'A}$ in 
 Eq.(\ref{F-eta-eta'}) is written in the following form,
\begin{gather}\label{Fee-abs}
 F_{\eta' \eta,B'A} = A^a_{\la \VV |}  \ov\sigma_{\eta'} |\ov S_{\eta',B'}\ra  \pi^a  |S_{\eta,A}\ra,
\end{gather}
 which can be directly verified by substituting Eqs.(\ref{pi-a}) and (\ref{sigma-eta})
 into the rhs of Eq.(\ref{Fee-abs}) and making use of Eqs.(\ref{Aa-Tmu}),
 (\ref{SA-SB-eta}), and (\ref{eps-delta}).
 Note that the term $A^a_{\la \VV |}$ does not change under LH spinor-space transformations.
 Although the expressions of the operators $\ov\sigma_\eta$ and $\pi^a$ given above include basis spinors,
  one can show that their forms are in fact basis-independent (see Appendix \ref{app-basis-ind}).
 Hence, these two operators should change in a direct way, that is,
\begin{gather}  \label{pia-ch}
 \pi^a \to \ww \pi^a = \sum_{\xi,\xi'} \tau^a_{\xi'\xi} |\ww S^A_{\xi'}\ra \la \ww S_{\xi,A}|,
 \\ \label{sigma-ch}
 \ov\sigma^\mu_\eta \to \ww{ \ov\sigma}^\mu_\xi =
 |T_\mu\ra \ov\sigma^{\mu}_{ D'C} \la \ww S^{C}_{\xi}| \la \ww{\ov S}^{D'}_{\xi}|,
\end{gather}
 where $\tau^a_{\xi'\xi} = \tau^a_{\eta'\eta}$ for $\xi=\eta$ and $\xi'=\eta'$.

 Then, substituting Eq.(\ref{|SAxi>}) and its complex conjugate and Eqs.(\ref{pia-ch})-(\ref{sigma-ch})
 into Eq.(\ref{Fee-abs}), one finds that
\begin{gather}\label{}
 F_{\eta' \eta, B'A} \to
  \sum_{\xi, \xi'} R_{\eta'\xi'} \ww F_{\xi' \xi} R^\dag_{\xi\eta}, \label{Feta-change}
\end{gather}
 where $\ww F_{\xi' \xi,B'A} =\ov\sigma^{\mu}_{B'A}  A_\mu^a \tau^a_{\xi'\xi}$.
 From Eqs.(\ref{F-eta-eta'}) and (\ref{Feta-change}),
 one finds the following change of $A_\mu^a \tau^a$,
\begin{gather}\label{Amua-change}
 A_\mu^a \tau^a_{\eta'\eta} \to \sum_{\xi, \xi'} R_{\eta'\xi'}
 (A_\mu^a \tau^a_{\xi'\xi}) R^\dag_{\xi\eta}.
\end{gather}
 It has the same formal form as the $SU(2)$ gauge transformation in Eq.(\ref{Aamu-gt-v}).

\subsection{Changes of the vacuum vector fields and the Lagrangian}\label{sect-LH-Avs}

 To complete our discussions, let us discuss changes of 
 the vacuum vector fields $A^{\rm vs }_{L\eta\eta', \mu}$ under
 LH spinor-space transformations.
 Substituting the transformations in Eqs.(\ref{psiL-change})
 and (\ref{phiL+-change}) into Eq.(\ref{op-Avs-L}), then, into Eq.(\ref{A-opA-L}), one finds that
\begin{gather}\label{Avs-LHST-change-1}
 A^{\rm vs }_{L\eta \eta' , \mu} \to  \N \sum_{\xi \xi'}  \left(\pp_\mu  R_{\eta\xi}  \la 0| \ww \psi_{L \xi }^A \right) 
 \ww \phi_{L \xi',A}^{\dag }  |0\ra R_{\xi'\eta'}^\dag.
\end{gather}
 Noting that Eqs.(\ref{Avs-Lambda-L}) and (\ref{<0psi-phi0>}) are also valid for the label $\xi$, one 
 can compute the rhs of Eq.(\ref{Avs-LHST-change-1}) and find that
\begin{gather}\label{Ars-tra3}
 A^{\rm vs }_{L\eta\eta', \mu}  \to A^{\rm vs }_{L\xi\xi', \mu}   -i \sum_{\xi }  (\pp_\mu R_{\eta\xi} )  R_{\xi\eta'}^\dag.
\end{gather}
 Thus, the vacuum vector fields transform in the same formal way as 
 in Eq.(\ref{Avs-LH-gt})  under $SU(2)$ gauge transformations.

 To summarize, under LH spinor-space transformations, the leptonic fields and the connection
 fields change in the ways given in Eqs.(\ref{psiL-change}), (\ref{psiL+-change}), 
 (\ref{Amua-change}) with $B_\mu(x)$ unchanged, and (\ref{Ars-tra3}).
 These changes have the same formal forms as the $SU(2)$ gauge transformations
 given in Eqs.(\ref{EL-gt}), (\ref{EL-gt+}), (\ref{SU-gt-v}), and (\ref{Avs-LH-gt}), respectively.
 One may dirctly verify invariance of  the LH Lagrangian $ \LL_l^{\rm LH}(x)$ in Eq.(\ref{LL-LH})  by writing  it as
\begin{gather}\label{LL-LH-2}
 \LL_l^{\rm LH} = i E_L^{\dag B'} \ov\sigma^{\mu}_{B'A}
 \left( \pp_\mu  -i gA^{a}_\mu \tau^a  -i y g'B_\mu - A^{\rm vs}_{L,\mu} \right) E_{L}^{A},
\end{gather}
 where Eq.(\ref{gamma-mu}) has been used to get the following expression of 
 $\gamma_0 \gamma^\mu$,
\begin{gather}
 \gamma_0 \gamma^\mu =
 \left( \begin{array}{cc} \ov\sigma^{\mu}_{B'A} & 0 \\ 0 & \sigma^{\mu DC'} \end{array} \right).
  \label{gamma0-gamma-mu}
\end{gather}
 Therefore, the form of the electroweak Lagrangian remains invariant
 under LH spinor-space transformations and this symmetry has 
 the same formal form as the $SU(2)$ gauge symmetry.

\section{Conclusions and discussions}\label{sect-conclusion}

 In this paper, it is proposed that the total connection field in the electroweak theory
 may include vacuum vector fields, as contributions from vacuum fluctuations of the leptonic fields,
 which have no experimentally-observable effect.
 It has been found that, with the vacuum vector fields taken into account, 
 the form of the total Lagrangian is invariant
 under local phase transformations of the basis states for leptons,
 as well as, under certain local transformations of the LH spinor spaces for leptons.
 And, changes of the leptonic fields and of the connection fields under these transformations
 possess the same formal forms as $U(1)$ and $SU(2)$ gauge transformations, respectively.

 The above results suggest that the gauge transformations in the electroweak theory may be interpreted as
 originating from certain changes of the basis states employed in the description of leptonic states.
 In particular, in this interpretation, $SU(2)$ gauge transformations do not really mix LH electron
 states and LH electron neutrino states, but, they mix basis states of the mathematical spaces that
 are employed in the description of the LH states.
 This gives a simple physical interpretation to the gauge symmetries;
 that is, the physics should not depend on the concrete bases employed,
as long as the bases possess the necessary mathematical properties for descriptions.

 It is worth future investigation whether the approach adopted in this paper may be useful
 for further understanding of the $SU(3)$ gauge symmetry in quantum chromodynamics.
 But, a direct application of the present method would not work,
 because the $SU(3)$ gauge symmetry involves a color degree of freedom,
 which is usually regarded as being independent of the spin degree of freedom.

 \acknowledgements

 The author is grateful to Yan Gu for valuable discussions and suggestions.
 This work was partially supported by the National Natural Science Foundation of China under Grant
 Nos.~11275179, 11535011, and 11775210.

\appendix

\section{$SL(2,C)$ and Lorentz transformations}\label{sect-SL2C-transf}

 In this appendix, we recall the relation between $SL(2,C)$ transformations and
 Lorentz transformations given in the spinor theory
 \cite{Penrose-book,CM-book,Corson,pra16-commu,Kim-group}.
 Particularly, when $SL(2,C)$ transformations are carried out on a
 space $\WW$, the corresponding transformations on the space $\VV$ are Lorentz transformations.

 The group $SL(2,C)$ is composed of $2\times 2$ complex matrices with unit determinant, written as
\begin{equation}\label{h-AB}
  h^{A}_{\ \ B} = \left( \begin{array}{cc} a & b \\ c & d \end{array} \right)
  \quad \text{with} \  ad-bc=1.
\end{equation}
 Under a transformation given by $h^{A}_{\ \ B}$,
 a two-component spinor $\kappa^A$ is transformed to
\begin{equation}\label{}
  \ww \kappa^A = h^{A}_{\ \ B} \kappa^B.
\end{equation}
 In this appendix, we use a tilde to indicate the result of a $SL(2,C)$ transformation.

 It is straightforward to verify that $\epsilon^{AB}$ is
 invariant under $SL(2,C)$ transformations, that is,
 $\ww\epsilon^{AB}= h^{A}_{\ \ C} h^{B}_{\ \ D} \epsilon^{CD}$
 has the same matrix form as $\epsilon^{AB}$ in Eq.(\ref{epsilon}).
 Direct computation can verify the following relations,
\begin{eqnarray} \label{h-property-1}
  h^A_{\ \ B} h_{C}^{\ \ B} = h^B_{\ \ C} h_{B}^{\ \ A}  = -\epsilon_C^{\ \ A} = -\delta_C^{A}.
\\  h_{AD} h^A_{\ \ C}  = \epsilon_{DC}, \quad h^A_{\ \ B} h^{C B} = \epsilon^{AC}.
\label{h-property-2}
\end{eqnarray}
 It is not difficult to verify that the product $\chi_A \kappa^A$ is a scalar product,
 that is, $\ww\chi_A \ww\kappa^A = \chi_A \kappa^A$.

 When $\kappa^A$ is transformed by a matrix $h^{A}_{\ \ B}$,
 $\ov\kappa^{A'}$ is transformed by its complex-conjugate matrix, namely,
\begin{equation}\label{}
  \ww {\ov\kappa}^{A'} = \ov h^{A'}_{\ \ B'} \ov\kappa^{B'},
\end{equation}
 where
\begin{equation}\label{}
 \ov h^{A'}_{\ \ B'} := (h^{A}_{\ \ B})^* \quad \text{with} \ A=A', B= B'.
\end{equation}

 Now, we discuss relation between $SL(2,C)$ transformations and Lorentz transformations.
 Related to a $SL(2,C)$ transformation $h^{A}_{\ \ B}$ performed on a space $\WW$,
 we use $\Lambda^\mu_{\ \ \nu}$ to denote the corresponding transformation on the space $\VV$,
\begin{equation}\label{ww-K}
   \ \ww K^\mu = \Lambda^\mu_{\ \ \nu} K^\nu .
\end{equation}
 It proves convenient to require invariance of the EM-symbols under $SL(2,C)$ transformations,
 namely,
\begin{equation}\label{wwsig=sig}
 \ww \sigma^{\mu A'B} = \sigma^{\mu A'B},
\end{equation}
 where
\begin{gather}\label{ww-sig}
 \ww \sigma^{\mu A'B} = \Lambda^\mu_{\ \ \nu} \ov h^{A'}_{\ \ C'}
 h^{B}_{\ \ D} \sigma^{\nu C'D}.
\end{gather}
 This requirement can fix the form of $\Lambda^\mu_{\ \ \nu}$.
 In fact, substituting Eq.(\ref{ww-sig})
 into Eq.(\ref{wwsig=sig}) and rearranging the positions of some labels, one gets
\begin{gather}\label{sigma-Lam-int1}
 \sigma^{\mu}_{\ A'B} = \Lambda^\mu_{\ \ \nu} \ov h_{A' C'} h_{B D} \sigma^{\nu C'D}.
\end{gather}
 Multiplying both sides of Eq.(\ref{sigma-Lam-int1})
 by $ \ov h^{A'}_{\ \ E'}h^B_{\ \ F} \sigma_{\nu}^{E'F}$, the rhs gives
\begin{gather}\label{}\notag
 \Lambda^\mu_{\ \ \eta} \ov h_{A' E'} h_{B F} \sigma^{\eta E'F}
  \ov h^{A'}_{\ \ C'} h^B_{\ \ D} \sigma_{\nu}^{C'D}
  \\ = \Lambda^\mu_{\ \ \eta} \epsilon_{E'C'} \epsilon_{FD} \sigma^{\eta E'F} \sigma_{\nu}^{C'D}
  = \Lambda^\mu_{\ \ \nu},
\end{gather}
 where Eq.(\ref{h-property-2}) and Eq.(\ref{st-delta}) have been used.
 Then, one gets the following expression for $\Lambda^\mu_{\ \ \nu}$,
\begin{equation}\label{Lam-s-h}
  \Lambda^\mu_{\ \ \nu} = \sigma^{\mu}_{A'B}  \ov h^{A'}_{\ \ C'} h^B_{\ \ D} \sigma_{\nu}^{C'D}.
\end{equation}

 Substituting  Eq.(\ref{Lam-s-h}) into the product
 $\Lambda^\mu_{\ \ \eta} \Lambda^\nu_{\ \ \xi} g^{\eta\xi}$, one gets
\begin{gather*}
 \sigma^{\mu}_{A'B}  \ov h^{A'}_{\ \ C'} h^B_{\ \ D} \sigma_{\eta}^{C'D}
 \sigma^{\nu}_{E'F}  \ov h^{E'}_{\ \ G'} h^F_{\ \ H} \sigma_{\xi}^{G'H} g^{\eta\xi}.
\end{gather*}
 Using Eq.(\ref{ss-ee-2}), this gives
\begin{gather*}
 \sigma^{\mu}_{A'B}  \ov h^{A'}_{\ \ C'} h^B_{\ \ D}
 \sigma^{\nu}_{E'F}  \ov h^{E'C'} h^{FD}.
\end{gather*}
 Then, noting Eqs.(\ref{h-property-2}) and (\ref{g-sig}), one gets the first equality in
 the following relations,
\begin{gather}\label{LLg=g}
 \Lambda^\mu_{\ \ \eta} \Lambda^\nu_{\ \ \xi} g^{\eta\xi} = g^{\mu\nu},
 \quad \Lambda^\mu_{\ \ \eta} \Lambda^\nu_{\ \ \xi} g_{\mu \nu} = g_{\eta \xi}.
\end{gather}
 The second equality in (\ref{LLg=g}) can be proved in a similar way.
 Therefore, the transformations $\Lambda^\mu_{\ \ \nu}$ constitute the
 (restricted) Lorentz group
 and the space $\VV$ is composed of four-component vectors.

 The transformations $\Lambda$ and the matrix $g$ have the following properties.
 (i) The inverse transformation of $\Lambda^\mu_{\ \ \nu}$, denoted by $\Lambda^{-1}$
 has the simple expression,
\begin{equation}\label{Lambda-1}
 (\Lambda^{-1})^\nu_{\ \ \mu} = \Lambda_\mu^{\ \ \nu} \Longleftrightarrow
 (\Lambda^{-1})_{\nu \mu} = \Lambda_{\mu \nu}.
\end{equation}
 In fact, substituting Eq.(\ref{Lam-s-h}) into the product
 $\Lambda^\mu_{\ \ \nu} \Lambda_\lambda^{\ \ \nu}$ and making use of
 Eqs.(\ref{h-property-1}), (\ref{st-delta}), and (\ref{f-AB}), it is straightforward
 to verify Eq.(\ref{Lambda-1}).

 (ii) Equation (\ref{LLg=g}) implies that the matrix
 $g^{\mu\nu}$ is invariant under the transformation $\Lambda$, that is,
\begin{equation}\label{wwg=g}
  \ww g^{\mu\nu} = g^{\mu\nu}.
\end{equation}

 (iii) The product $K^\mu g_{\mu\nu} J^\nu = K_\mu J^\mu$ is a scalar under the
 transformation $\Lambda$, i.e.,
\begin{equation}\label{<K|J>}
 \ww K_\mu \ww J^\mu = K_\mu J^\mu,
\end{equation}
 which can be readily proved making use of Eq.(\ref{LLg=g}).

 (iv) Making use of Eq.(\ref{sig-c}), it is straightforward to show that
the transformation $\Lambda $ is real, namely,
\begin{equation}\label{real-Lam}
 \Lambda^\mu_{\ \ \nu} = (\Lambda^\mu_{\ \ \nu})^*.
\end{equation}
 Then, it is easy to check that $K^*_\mu J^\mu$ is also a scalar product.

\section{Basis-independence of $\ov\sigma_\eta$ and $\pi^a$}\label{app-basis-ind}

 In this appendix, we show that  the forms of the operators $\ov\sigma_\eta$
 and $\pi^a$ are basis-independent.
 To this end, one may consider an arbitrary $SL(2,C)$ transformation $h^A_{\ \ B}$,
 which connects $|S_{\eta}^A\ra$ and $|S^A_{\eta'}\ra $ to new basis states written as
 $|s_{\eta}^A\ra$ and $|s^A_{\eta'}\ra $, respectively, that is,
\begin{subequations}
\begin{gather}\label{S-h-s}
 |S_{\eta}^A\ra  = h^A_{\ \ B} |s_{\eta}^B\ra,
 \\ |S^A_{\eta'}\ra =   h^A_{\ \ B} |s^B_{\eta'}\ra. \label{S-h-s'}
\end{gather}
\end{subequations}

 Let us first discuss the operator $\ov\sigma_\eta$ in Eq.(\ref{sigma-eta}).
 Under the $SL(2,C)$ transformation in Eq.(\ref{S-h-s}),
 basis bras in the spaces dual to $\WW_\eta$ and $\ov\WW_\eta$ change as follows,
\begin{gather}\label{<S-h-s|}
 \la S_{\eta}^A|  = h^A_{\ \ B} \la s_{\eta}^B|, \quad
 \la \ov S_{\eta}^{A'}|  = \ov h^{A'}_{\ \ B'} \la s_{\eta}^{B'}|.
\end{gather}
 Meanwhile, a Lorentz transformation related to $h^A_{\ \ B}$, denoted by $\Lambda^\mu_{\  \nu}$,
 should connect the basis states $|T_\mu\ra$
 to a new set of basis vectors, denoted by $|t_\mu\ra$, that is,
\begin{gather}\label{T-t}
 |T_\mu\ra = \Lambda_\mu^{\  \nu} |t_\nu\ra.
\end{gather}
 Substituting Eqs.(\ref{<S-h-s|})-(\ref{T-t}) into Eq.(\ref{sigma-eta})
 and making use of the invariance of the EW-symbols under $SL(2,C)$ transformations
 shown in Eq.(\ref{wwsig=sig}), one finds that
\begin{gather}\label{sigma-eta-st}
 \ov\sigma_\eta = |t_\mu\ra \ov\sigma^{\mu}_{ D'C} \la s^{C}_{\eta}| \la \ov s^{D'}_{\eta}|.
\end{gather}
 Hence, the form of  $\ov\sigma_\eta$ is basis-independent.

 Next, we discuss $\pi^a$.
 Noting the relations in Eqs.(\ref{<kappa|}) and (\ref{f-AB}), from Eq.(\ref{S-h-s}) one gets that
 $\la S_{\eta,A}|  = -h_{A}^{\ \ B} \la s_{\eta, B}|$.
 Substituting this result and Eq.(\ref{S-h-s'}) into Eq.(\ref{pi-a}), one finds that
\begin{gather}\notag 
 \pi^a = -\sum_{\eta,\eta'} \tau^a_{\eta'\eta} h^A_{\ \ B} |s_{\eta'}^B\ra h_{A}^{\ \ C} \la s_{\eta, C}|.
\end{gather}
 Then, using Eq.(\ref{h-property-1}), one gets that
\begin{gather}\label{pi-a-s}
 \pi^a = \sum_{\eta,\eta'} \tau^a_{\eta'\eta} |s_{\eta'}^B\ra \la s_{\eta, B}|,
\end{gather}
 showing that the form of  $\pi^a$ is also basis-independent.



\begin{thebibliography}{99}
\bibitem{footnote-prefactor}
 The field $\psi_\eta$ has some constant prefactor, which we do not write explicitly for brevity.
 In the literature, the factor $\frac{1}{p^0}$ in $d\ww p$ is sometimes written as
 $\frac{1}{\sqrt{p^0}}$.
 Here, we write this form of $d\ww p$, because it is Lorentz-invariant.
 Consistently, the anti-commutation relation for creation and annihilation operators
 has a factor $p^0$ [cf. Eq.(\ref{b-bdag=0})].
\bibitem{footnote-Avsmu} For the  operator $\A^{\rm vs }_{L\eta,\mu}$, we do not
 consider the possibility of  $\psi_{L \eta}^{\dag B'} (\pp_\mu \psi_{L \eta }^A) $,
 because its dependence on the two labels $A$ and $B'$ can not be taken out in a simple and Lorentz covariant way.
\end{thebibliography}
\end{document}